\documentclass{jfm}
\usepackage{graphicx}
\usepackage{epstopdf, epsfig}
\usepackage{amsmath}
\usepackage{commath}

\usepackage[section]{placeins}
\usepackage{caption}
\usepackage{subcaption}
\usepackage{xcolor}
\usepackage{float}

\title{The rotation of a sedimenting anisotropic particle in a linearly stratified ambient}
\shorttitle{Torque on an anisotropic particle in a linearly stratified ambient}
\shortauthor{G. Subramanian}

\author{Arun Kumar Varanasi\aff{1}
  \corresp{\email{arun.target@gmail.com}},
  Navaneeth K. Marath\corresp{\email{navaneeth@iitrpr.ac.in}}\aff{2},
 \and Ganesh Subramanian\corresp{\email{sganesh@jncasr.ac.in}}\aff{1}}

\affiliation{\aff{1}Engineering Mechanics Unit, Jawaharlal Nehru Center for Advanced Scientific Research, Bangalore-560064, India
\aff{2}Mechanical Engineering,Indian Institute of Technology, Ropar 140001 ,India}

\begin{document}

\maketitle

\begin{abstract}
We derive analytically the torque on a spheroid of an arbitrary aspect ratio $\kappa$, sedimenting in a linearly stratified ambient. The analysis demarcates regions in parameter space corresponding to broadside-on and edgewise (longside-on) settling in the limit $Re, Ri_v \ll 1$, where $Re = \rho_0UL/\mu$ and $Ri_v =\gamma L^3g/\mu U$, the Reynolds and viscous Richardson numbers, respectively, are dimensionless measures of the importance of inertial and buoyancy forces relative to viscous ones. Here, $L$ is the spheroid semi-major axis, $U$ an appropriate settling velocity scale, $\mu$ the fluid viscosity, and $\gamma\,(>0)$ the (constant)\,density gradient characterizing the stably stratified ambient, with $\rho_0$ being the fluid density taken to be a constant within the Boussinesq framework. A generalized reciprocal theorem formulation identifies three contributions to the torque: (1) an $O(Re)$ inertial contribution that already exists in a homogeneous ambient, and orients the spheroid broadside-on; (2) an $O(Ri_v)$ hydrostatic contribution due to the ambient linear stratification that also orients the spheroid broadside-on; and (3) a hydrodynamic contribution arising from the perturbation of the ambient stratification by the spheroid whose nature depends on $Pe$; $Pe = UL/D$ being the Peclet number with $D$ the diffusivity of the stratifying agent. For $Pe \ll 1$, this contribution is $O(Ri_v)$ and orients prolate spheroids edgewise for all $\kappa\,(>1)$. For oblate spheroids, it changes sign across a critical aspect ratio $\kappa_c \approx 0.17$, orienting oblate spheroids with $\kappa_c < \kappa < 1$ edgewise and those with $\kappa < \kappa_c$ broadside-on. For $Pe \ll 1$, the hydrodynamic component is always smaller in magnitude than the hydrostatic one, so a sedimenting spheroid in this limit always orients broadside-on. In contrast, for $Pe \gg 1$, the hydrodynamic contribution is dominant, being $O(Ri_v^{\frac{2}{3}}$) in the Stokes stratification regime characterized by $Re \ll Ri_v^{\frac{1}{3}}$, and orients the spheroid edgewise regardless of $\kappa$. The differing orientation dependencies of the inertial and large-$Pe$ hydrodynamic stratification torques imply that the broadside-on and edgewise settling regimes are separated by two distinct $\kappa$-dependent critical curves in the $Ri_v/Re^{\frac{3}{2}}-\kappa$ plane, with the region betwen these curves correponding to stable intermediate equilibrium orientations. The predictions for large $Pe$ are broadly consistent with recent experimental observations.
\end{abstract}

\section{Introduction}
The atmosphere and oceans being, on average, in a stably stratified state, the motion of particles as well as living organisms in a stratified ambient is of obvious importance in natural settings. A large fraction of the research on the vertical motion of particles through stratified fluids, including cases of both sharp\,(\cite{Camassa_2009}\cite{Fernando_1999}) and continuous\,(\cite{Hanazaki_1988},\cite{Hanazaki_2009},\cite{yick_2009},\cite{doostmohammadi_2012},\cite{doostmohammadi_2014}) stratification profiles, has, however, focused on spherical particles. Although this research has shed light on the non-trivial effects of stratification on the structure of the disturbance flow field induced by a sedimenting sphere, for instance, its sensitive dependence on the diffusivity of the stratifying agent via the Peclet number\,($Pe$)\,(for instance, see \cite{list_laminar_1971}, \cite{ardekani_2010}, \cite{doostmohammadi_2012} and \cite{Arun_2020}), the vast majority of particles and living (micro)organisms in natural scenarios depart from the idealized spherical shape. Indeed, both marine phytoplankton and zooplankton come in an astonishing variety of shapes\,(\cite{PhytoGuide2018},\cite{Thomas_2011}), and there are provocative questions to be addressed with regard to the large-scale effects of zooplankton migration across the oceanic pycnocline\,(\cite{kunze_2006},\cite{visser_2007},\cite{katija_2009},\cite{subramanian_2010},\cite{Arun_2020}). Other classes of organic particles including marine snow aggregates\,(\cite{Prairie_2015}), phytodetritus and faecal pellets, which make up the so-called biological pump\,(\cite{Turner_2015}), and undesired microplastics\,(\cite{Turner_2011},\cite{Cole_2011}), also depart significantly from the canonical spherical geometry. Extensive research over a long time has now led to a fairly mature understanding of the dynamics of anisotropic particles sedimenting in a homogeneous ambient\,(\cite{Magnaudet_2002},\cite{Magnaudet_2013}). While the non-trivial effects of unsteady wake dynamics come into play at higher Reynolds numbers\,($Re$), as manifest by the onset of path instabilities of sedimenting spheroids\,(\cite{Magnaudet_2012}), the simplest scenario which prevails for low to moderate Reynolds numbers, when the wake has a quasi-steady character, involves inertial effects acting to turn sedimenting anisotropic particles broadside-on. For small $Re$, and in the case where the anisotropic particle is a prolate or an oblate spheroid, the inertial torque acting to turn the spheroid broadside-on has been determined analytically as a function of the spheroid aspect ratio\,(\cite{Marath_JFM2015}), and has recently been shown to exert a crucial influence on the orientation distribution of such particles settling through a turbulent velocity field\,(\cite{Prateek_2020}).

The present effort is specifically motivated by very recent experiments involving cylindrical and disk-shaped particles (\cite{mercier_2020},\cite{Mrokowska_2018},\cite{Mrokowska_20201},\cite{Mrokowska_20202}) that are among the first to systematically explore the role of shape anisotropy for sedimenting particles in a heterogeneous stably stratified ambient. The experiments reported by \cite{mercier_2020} pertain to a linearly stratified ambient, while those reported in \cite{Mrokowska_2018},\cite{Mrokowska_20201} and \cite{Mrokowska_20202} pertain to a non-linearly stratified fluid layer sandwiched between homogeneous upper and lower layers. While the detailed results obtained for the two sets of experiments differ on account of the differing nature of the ambient stratification, one of the most important findings, common to both sets of experiments, pertains to the ability of the torque due to buoyancy forces to oppose, and even overwhelm the aforementioned inertial torque that acts in a homogeneous setting, thereby turning the particle longside-on. A recent theoretical effort\,(\cite{dandekar_2020}) has attempted to calculate the stratification-induced corrections to the force and torque acting on a non-spherical particle settling in a viscous linearly stratified ambient. While a correction to the force was determined in terms of the viscous Richardson number\,($Ri_v$ defined below in section \ref{sec:recithm}) for both chiral and achiral particles, a hydrodynamic torque was found to arise from buoyancy forces only for chiral particles, the origin of this torque being the translation-rotation coupling that already exists for such particles in a homogeneous ambient. Thus, the analysis does not explain the principal observation in the aforementioned experiments involving the stratification-induced transition of a sedimenting anisotropic but achiral particle from a broadside-on to an edgewise configuration. In the present effort, we show that buoyancy forces associated with the ambient stratification do lead to a torque  for achiral particles modeled as prolate and oblate spheroids of an arbitrary aspect ratio. This stratification-induced torque consists of both hydrostatic and hydrodynamic components, with only the former contribution having been calculated in \cite{dandekar_2020} for a few restricted particle geometries. We show that the hydrostatic contribution always turns a spheroid broadside-on, regardless of aspect ratio, in agreement with \cite{dandekar_2020}. More importantly, the hydrodynamic component of the stratification-induced torque is shown to be asymptotically larger than the hydrostatic one for large $Pe$, and orients spheroids edgewise, thereby offering the first explanation of the experimental observations above, of edgewise settling of an anisotropic particle in a stratified fluid.

The layout of the paper as follows. In section \ref{sec:recithm}, we describe the reciprocal theorem formulation which yields the torque acting on a spheroid sedimenting in a linearly stratified viscous ambient in terms of distinct contributions arisng from the effects of fluid inertia and the buoyancy forces associated with the ambient stratification. The fluid inertial torque and the hydrodystatic component of the stratification torque are readily evaluated on account of their regular character, and this calculation is given in section \ref{sec:inertiastatic:torque}. The calculation of the hydrodynamic component of the stratification torque is more involved, being dependent on $Pe$, and is carried out in sections \ref{sec:smallPe} and \ref{sec:largePe} in the limits $Pe \ll 1$ and $Pe \gg 1$, respectively. Finally, section \ref{sec:results} discusses the transition from broadside-on to edgewise settling that arises due to the completing influences of the inertial and hydrodynamic components of the stratification torque, at large $Pe$, and ends with a comparison with recent experiments. In section \ref{sec:conclusions}, we briefly indicate possible lines of investigation for the future.

\section{A sedimenting spheroid in a linearly stratified ambient: The generalized Reciprocal theorem formulation} \label{sec:recithm}

The torque acting on a spheroid, sedimenting in a stably stratified ambient, is derived below using the generalized reciprocal theorem\,(see \cite{KimKarrila1991};\cite{Marath_JFM2015,Marath_JFM2016}). The theorem relates two pairs of stress and velocity fields, and may be stated in the form:
\begin{align}
\displaystyle\int_{S_p} \sigma^{(2)}_{ij} u^{(1)}_i n_j dS - \displaystyle\int_{S_p} \sigma^{(1d)}_{ij} u^{(2)}_i n_j dS =&\,\displaystyle\int \frac{\partial \sigma_{ij}^{(1d)}}{\partial x_j} u^{(2)}_i dV, \label{gen_recithm:1}
\end{align}
where $S_p$ denotes the surface of the spheroid, and with the neglect of the surface integral at infinity, the volume integral on the right hand side of (\ref{gen_recithm:1}) is over the unbounded fluid domain external to the spheroid. In (\ref{gen_recithm:1}), the pair $({\boldsymbol \sigma}^{(1d)},{\boldsymbol u}^{(1)})$ denotes the dynamic stress and velocity fields associated with the problem of interest viz.\! a torque-free spheroid sedimenting under gravity in an ambient linearly stratified medium for small Reynolds\,($Re$) and viscous Richardson\,($Ri_v$) numbers. These dimensionless parameters measure the relative importance of inertial and buoyancy forces relative to viscous forces, respectively, and the aforementioned limit corresponds to the case where inertia and stratification act as weak perturbing influences about a leading order Stokesian approximation. The Reynolds and Richardson numbers are defined later in this section when writing down the non-dimensional system of governing equations; the precise definition of the dynamic stress field, ${\boldsymbol \sigma}^{(1d)}$, is also provided at the same place. The pair $({\boldsymbol \sigma}^{(2)},{\boldsymbol u}^{(2)})$, that defines the test problem in (\ref{gen_recithm:1}), corresponds to the stress and velocity fields associated with the Stokesian rotation of the same spheroid, about an axis orthogonal to its axis of symmetry, in a homogeneous and otherwise quiescent ambient with the same (assumed constant) viscosity as the medium in the actual problem. The equations governing the test problem may be written as:
\begin{align}
\frac{\partial u^{(2)}_i}{\partial x_i} =&\,0,  \label{test:eqcont} \\
\mu \frac{\partial^2 u^{(2)}_{i}}{\partial x_j^2} - \frac{\partial p^{(2)}}{\partial x_i} =&\,0, \label{test:eqmotion}
\end{align}
with the boundary condition ${\boldsymbol u}^{(2)} = {\boldsymbol \Omega}^{(2)} \wedge {\boldsymbol x}$ on $S_p$, ${\boldsymbol \Omega}^{(2)}$ being the angular velocity of the spheroid in the test problem, and the far-field decay condition for both ${\boldsymbol u}^{(2)}$ and $p^{(2)}$. Use of this surface boundary condition in (\ref{gen_recithm:1}) leads to 
\begin{align}
\displaystyle\int_{S_p} \sigma^{(2)}_{ij} u^{(1)}_i n_j dS - \Omega^{(2)}_j \displaystyle\int_{S_p} \epsilon_{ijk} x_k \sigma^{(1d)}_{il}  n_l dS =&\,\displaystyle\int \frac{\partial \sigma^{(1d)}_{ij}}{\partial x_j} u^{(2)}_i dV \label{gen_recithm:2}
\end{align}
where the second integral on the left hand side in (\ref{gen_recithm:2}) now denotes the torque due to the dynamic stress field ${\boldsymbol \sigma}^{(1d)}$. We postpone further simplification of (\ref{gen_recithm:2}) until after we define the pair $({\boldsymbol \sigma}^{(1d)},{\boldsymbol u}^{(1)})$ below.

As mentioned above, problem 1 corresponds to an arbitrarily oriented spheroid, sedimenting under the action of a gravitational force $F\hat{\boldsymbol g}$, in an ambient medium that is linearly stratified\,(along $\hat{\boldsymbol g}$) in the absence of the fluid motion induced by the spheroid. The unit vector $\hat{\boldsymbol g}$ is aligned along gravity, with $g$ denoting the magnitude of the gravitational acceleration, and $F = \frac{4\pi}{3}Lb^2\Delta \rho g\,(\frac{4\pi}{3}L^2b \Delta \rho g)$ denoting the buoyant weight for a prolate\,(oblate) spheroid. Here, $L$ and $b$ are the semi-major and semi-minor axes of the spheroid, with $\kappa = L/b$ being the aspect ratio of a prolate spheroid, and $\kappa = b/L$ that of an oblate spheroid; thus, $\kappa >1$ and $<1$ for prolate and oblate spheroids, respectively. The density difference that enters the buoyant weight above is $\Delta \rho = \rho_s -\rho^{(1)}_\infty({\boldsymbol x}_c)$, with $\rho_s$ being the density of the spheroid\,(assumed homogeneous), and $\rho^{(1)}_\infty({\boldsymbol x}_c) = \rho_0$ being the ambient fluid density at the center of the spheroid. The latter simplification arises because of the linear stratification and the fore-aft symmetry of the spheroid, both of which imply that the weight of the equivalent stratified spheroidal fluid blob that gives the buoyant force is the same as the weight of a homogeneous fluid blob with density equal to the ambient value at the spheroid center. In a lab-fixed reference frame, the ambient density field in problem 1 may be written in the form:
\begin{align}
\rho^{(1)}_\infty({\boldsymbol x}^L) =\rho_0 + \gamma x^L_i \hat{g}_i
\end{align}
where ${\boldsymbol x}^L$ denotes the position vector in laboratory coordinates with the spheroid center as the origin, and $\gamma > 0$ is the constant density gradient that characterizes the stable ambient stratification. The calculations for the torque are, however, best done in a reference frame translating with the spheroid where a quasi-steady state is assumed to prevail at leading order. The latter assumption is motivated by the asymptotically weak rotation of the sedimenting spheroid in the limit $Re, Ri_v \ll1$. The precise condition for the quasi-steady state assumption to hold depends on $Pe$, being more restrictive for large $Pe$, and is stated later alongside the results for the spheroid angular velocity for small and large $Pe$, obtained below. 

The ambient density in the particle-fixed reference frame takes the form:
\begin{align}
\rho^{(1)}_\infty({\boldsymbol x}) =\rho_0 +\gamma ( x_i + U_i t)\hat{g}_i,  \label{ambdens:particlefixed}
\end{align}
${\boldsymbol x}$ being the position vector in the new reference frame. In (\ref{ambdens:particlefixed}), ${\boldsymbol U}$ is the spheroid settling velocity, and related to the force\,($F\hat{\boldsymbol g}$) via a mobility tensor that is a known function of the spheroid aspect ratio $\kappa$. In terms of the spheroid orientation vector ${\boldsymbol p}$, one may write ${\boldsymbol U} = \frac{1}{\mu L} [X_A^{-1}{\boldsymbol p}{\boldsymbol p}+ Y_A^{-1}({\boldsymbol I} - {\boldsymbol p}{\boldsymbol p})] \cdot (F\hat{\boldsymbol g})$, $X_A(\kappa)$ and $Y_A(\kappa)$ being the axial and transverse translational resistance functions. The aspect ratio dependence of these functions is well known\,(see \cite{KimKarrila1991}), and is given in Appendix \ref{appendixresistance} for convenient reference. Note that the ambient density at the center of the spheroid\,(${\boldsymbol x} = 0$) is given by $\rho_0 + \gamma(U_i\hat{g}_i)t$, the time dependence arising from the spheroid translation. The equations of motion for problem 1, within a Boussinesq framework where the fluid density multiplying the inertial terms is taken as a constant $\rho_0$\,(say), may be written as:
\begin{align}
\frac{\partial u^{(1)}_i}{\partial x_i} =&\,0, \label{prob1:eqcont} \\
\mu \frac{\partial^2 u^{(1)}_{i}}{\partial x_j^2} - \frac{\partial p^{(1)}}{\partial x_i} =&\, \rho_0 u^{(1)}_j\frac{\partial u^{(1)}_i}{\partial x_j} -  \rho^{(1)}g_i,
 \label{prob1:eqmot}\\
\frac{\partial \rho^{(1)}}{\partial t} + u^{(1)}_j\frac{\partial \rho^{(1)}}{\partial x_j} =&\,D \nabla^2 \rho^{(1)}. \label{prob1:densitycons}
\end{align}
One now defines the perturbation density via $\rho^{(1)} = \rho_0 + \gamma (x_i + U_i t)\hat{g}_i + \rho'^{(1)}$. Next, using the scales $U=F/(\mu L X_A)$ for the velocity, $L$ for the length, $\mu U/L$ for the pressure and $\gamma L$ for $\rho'^{(1)}$, one obtains the following system of non-dimensional equations:
\begin{align}
\frac{\partial u^{(1)}_i}{\partial x_i} =&\,0, \label{prob1:eqcontND} \\
\frac{\partial^2 u^{(1)}_{i}}{\partial x_j^2} - \frac{\partial p^{(1)}}{\partial x_i} + \frac{\rho_0 g L^2}{\mu U}\hat{g}_i + Ri_v (\hat{U}_j  t +x_j )\hat{g}_j\hat{g}_i=&\, Re \,u^{(1)}_j\frac{\partial u^{(1)}_i}{\partial x_j} -  Ri_v\rho'^{(1)}\hat{g}_i,
 \label{prob1:eqmotND}\\
u^{(1)}_j\frac{\partial \rho'^{(1)}}{\partial x_j} + (\hat{U}_j + u^{(1)}_j)\hat{g}_j =&\,\frac{1}{Pe} \nabla^2 \rho'^{(1)}. \label{prob1:densityconsND}
\end{align}
where $Re = \rho_0UL/\mu$ and $Ri_v = \gamma L^3g/(\mu U)$ are the Reynolds and viscous Richardson numbers; we continue to use the same notation for the dimensionless fields for simplicity, with $\hat{\boldsymbol U}$ now being a dimensionless vector along the direction of settling; note that $\hat{\boldsymbol U}$ is not a unit vector for an arbitrarily oriented spheroid, and reduces to one only for a spheroid aligned with gravity. In the particle-fixed frame, ${\boldsymbol u}^{(1)} \rightarrow -\hat{\boldsymbol U}$ at large distances, corresponding to the ambient uniform flow far away from the particle, and thus, the combination $(\hat{\boldsymbol U} + {\boldsymbol u}^{(1)}) \cdot \hat{\boldsymbol g}$ in (\ref{prob1:densityconsND}) denotes the convection of the (constant)\,base-state density gradient by the component of the disturbance velocity field along gravity. Finally, the time dependence of ${\boldsymbol u}^{(1)}$ in (\ref{prob1:eqmotND}), and that of $\rho'^{(1)}$ in (\ref{prob1:densityconsND}) in particular, that arise from the (slow)\,rotation of the spheroid, have been neglected owing to the quasi-steady state assumption made in (\ref{prob1:eqcontND}-\ref{prob1:densityconsND}); the time dependence of the density multiplying the inertial terms, on account of spheroid translation, has already been neglected within the Boussinesq approximation. 

One now defines a disturbance pressure field via $p^{(1)} = p_0^{(1)} + p'^{(1)}$ with 
\begin{align}
\frac{\partial p_0^{(1)}}{\partial x_i} =&\frac{\rho_0 g L^2}{\mu U}\hat{g}_i + Ri_v (\hat{U}_j t +x_j) \hat{g}_j\hat{g}_i, \label{hydrostat:p}
\end{align}
so that $p^{(1)}_0$ defines the baseline hydrodystatic contribution arising from the ambient linear stratification. Having incorporated the baseline hydrostatic variation in $p^{(1)}_0$, and defining the disturbance velocity field as ${\boldsymbol u}^{(1)} = -\hat{\boldsymbol U} +{\boldsymbol u}'^{(1)}$, one may write the governing equations above in terms of the disturbance velocity, pressure and density fields as follows:
\begin{align}
\frac{\partial u'^{(1)}_i}{\partial x_i} =&\,0, \label{phys:eqcont} \\
\frac{\partial \sigma^{(1d)}_{ij}}{\partial x_j} =&\, Re (-\hat{U}_j+ u'^{(1)}_j)\frac{\partial u'^{(1)}_i}{\partial x_j} -  Ri_v  \rho'^{(1)}\hat{g}_i, \label{phys:eqmot}\\
(-\hat{U}_j + u'^{(1)}_j)\frac{\partial \rho'^{(1)}}{\partial x_j} =&\,  -u'^{(1)}_j\hat{g}_j + \frac{1}{Pe} \nabla^2\rho'^{(1)}. \label{phys:densitycons}
\end{align}
where the left hand side of (\ref{prob1:eqmotND}) has been written in terms of the dynamic stress field ${\boldsymbol \sigma}^{(1d)}$ defined by ${\boldsymbol \sigma}^{(1d)} = -p'^{(1)}{\boldsymbol I} + ( {\boldsymbol \nabla} {\boldsymbol u}'^{(1)} + {\boldsymbol \nabla} {{\boldsymbol u}'^{(1)}}^\dag)$. Thus, one has the relation ${\boldsymbol \sigma}^{(1)} = -p_0^{(1)} {\boldsymbol I} + {\boldsymbol \sigma}^{(1d)}$ between the total and the dynamic stress fields of problem 1.

Assuming the spheroid in problem 1 to rotate with an angular velocity ${\boldsymbol \Omega}^{(1)}$, one has ${\boldsymbol u}^{(1)} = {\boldsymbol \Omega}^{(1)} \wedge {\boldsymbol x}$ on $S_p$. Using this in the first surface integral in (\ref{gen_recithm:1}), and substituting the divergence of the dynamic stress from (\ref{phys:eqmot}) in the volume integral in (\ref{gen_recithm:1}), one obtains:
\begin{align}
\Omega^{(1)}_j  {\mathcal L}_j^{(2)} -\Omega^{(2)}_j {\mathcal L}_j^{\sigma(1)d} =&\,Re\displaystyle\int u^{(2)}_i \, (-\hat{U}_j+ u'^{(1)}_j)\frac{\partial u'^{(1)}_i}{\partial x_j} dV - Ri_v   \displaystyle\int \rho'^{(1)} \hat{g}_i\, u_{i}^{(2)}\,dV
\end{align}
where $\boldsymbol{\mathcal L}^{\sigma(1)d}$ now denotes the torque contribution due to the dynamic stress ${\boldsymbol \sigma}^{(1d)}$.
Now, the particle in problem 1 is torque-free. In light of the above relation between ${\boldsymbol \sigma}^{(1)}$ and ${\boldsymbol \sigma}^{(1d)}$, the  total torque may be written as $\boldsymbol{\mathcal L}^{(1)} = \boldsymbol{\mathcal L}^{\sigma(1)d} + \boldsymbol{\mathcal L}^{\sigma(1)s}$, where the dynamic torque component $\boldsymbol{\mathcal L}^{\sigma(1)d}$ includes both inertia and stratification-induced contributions, while $\boldsymbol{\mathcal L}^{\sigma(1)s}$ is the hydrostatic contribution due to the pressure field $p^{(1)}_0$ associated with the linearly varying density field of the stably stratified ambient, and defined by (\ref{hydrostat:p}). Thus, $\boldsymbol{\mathcal L}^{(1)} = 0 \Rightarrow\boldsymbol{\mathcal L}^{\sigma(1)d}=-\boldsymbol{\mathcal L}^{\sigma(1)s}$, and the relation involving the spheroid angular velocity in problem 1 takes the following form:
\begin{align}
 \Omega^{(1)}_j  {\mathcal L}_j^{(2)}= Re\int u_{i}^{(2)} \,(-\hat{U}_j+ u'^{(1)}_j)\frac{\partial{u'^{(1)}_i}}{\partial x_j} dV - \biggl[\Omega^{(2)}_j {\mathcal L}_j^{\sigma(1)s} + Ri_v \int \rho'^{(1)} \hat{g}_i\, u_{i}^{(2)}\,dV \biggr]. \label{recithm_Angvel}
\end{align}
where
\begin{align}
 {\mathcal L}_i^{\sigma(1)s} =&\, -\epsilon_{ijk} \displaystyle\int p_0^{(1)}  x_j n_k dS, \label{hydrostat:torque}
\end{align}
with $p_0^{(1)}$  being defined in (\ref{hydrostat:p}). Since the buoyancy force in a homogenous ambient acts through the centre of the spheroid, only the linearly varying term in (\ref{hydrostat:p}) contributes to the hydrostatic torque, which may therefore be written as:
\begin{align}
 {\mathcal L}_i^{\sigma(1)s} =&\, -Ri_v\, \epsilon_{ijk} \displaystyle\int \frac{1}{2}(x_l\hat{g}_l)^2x_j  n_k dS, \label{hydrostat:torque}
\end{align}
the contribution above remaining the same regardless of the choice of reference frame\,(${\boldsymbol x}$ or ${\boldsymbol x}^L$). On substitution of the above expression for ${\mathcal L}_k^{\sigma(1)s}$, and using the relation ${\boldsymbol u}^{(2)}_i = {\boldsymbol U}^{(2)}_{ij}{\boldsymbol \Omega}^{(2)}_j$ (on account of the linearity of the Stokes equations), the second order tensor ${\boldsymbol U}^{(2)}_{ij}$ being known in closed form\,(see \cite{Marath_JFM2015}, and section \ref{sec:inertiastatic:torque} below), (\ref{recithm_Angvel}) takes the form:
\begin{align}
 \Omega^{(1)}_j  {\mathcal L}_j^{(2)}= &\Omega^{(2)}_k\left\{ Re\int U_{jk}^{(2)} \, (-\hat{U}_l+ u'^{(1)}_l)\frac{\partial{u'^{(1)}_j}}{\partial x_l} dV \right.\nonumber \\
  &\left.  - Ri_v \biggl[-\frac{1}{2} \epsilon_{klm}  \displaystyle\int (x_j\hat{g}_j)^2 x_l  n_m dS+  \int \rho'^{(1)} \hat{g}_j\, U_{jk}^{(2)}\,dV \biggr] \right\}, \label{recithm_Angvelf}
\end{align}
Again, on account of linearity, one may write the torque on the rotating spheroid, in the test problem, in the form $\boldsymbol{\mathcal L}^{(2)} =[X_C {\boldsymbol p}{\boldsymbol p} + Y_C({\boldsymbol I} - {\boldsymbol p}{\boldsymbol p})] \cdot \boldsymbol{\Omega}^{(2)}$, where $X_C$ and $Y_C$ are the axial and transverse rotational resistance functions, and are known functions of $\kappa$\,(\cite{KimKarrila1991}), whose expressions are given in Appendix \ref{appendixresistance}. By symmetry, the sedimenting spheroid cannot spin about its axis regardless of its orientation, and therefore without loss of generality, the test problem 2 can be taken as that of a tranverse Stokesian rotation\,(${\boldsymbol \Omega}^{(2)} \cdot {\boldsymbol p} = 0$), in which case the test-torque-angular-velocity relation takes the simpler form $\boldsymbol{\mathcal L}^{(2)} = Y_C \boldsymbol{\Omega}^{(2)}$. Finally, accounting for the fact that the test angular velocity ${\boldsymbol \Omega}^{(2)}$ can point in an arbitrary direction in a plane perpendicular to ${\boldsymbol p}$, one obtains the following relation for the spheroid angular velocity in problem 1:
\begin{align}
 \Omega^{(1)}_i=&\, \frac{1}{Y_C}\left\{ Re\int U_{ji}^{(2)} \, (-\hat{U}_l+ u'^{(1)}_l)\frac{\partial{u'^{(1)}_j}}{\partial x_l} dV + Ri_v \biggl[ \epsilon_{ilm} \displaystyle\int\frac{1}{2} (x_j\hat{g}_j)^2 x_l n_m dS  \biggr. \right.. \nonumber \\
 &\hspace*{2.5in} -\left. \biggl. \displaystyle\int \rho'^{(1)} \hat{g}_j\, U_{ji}^{(2)}\,dV \biggr] \right\}, \label{recithm_Angvelfinal}
\end{align}
Since a settling spheroid in a homogeneous ambient must retain its initial orientation in the Stokes limit on account of reversibility, expectedly, the rotation of the spheroid, as given by (\ref{recithm_Angvelfinal}), arises due to the combined (weak)\,effects of fluid inertia and the ambient stratification. The first term in (\ref{recithm_Angvelfinal}) corresponds to the inertial torque, while the second and third terms which have been grouped together correspond to the hydrostatic and hydrodynamic components of the stratification torque, respectively. The hydrostatic torque only involves knowledge of the ambient density field, and is easily evaluated. The inertial torque has a regular character in that the dominant contributions to the $O(Re)$ volume integral in (\ref{recithm_Angvelfinal}) arise from a volume of $O(L^3)$ around the sedimenting spheroid, and therefore, the integral may again readily be determined at leading order using the Stokesian approximations for the velocity fields involved, as has been done in \cite{Marath_JFM2015}. The evaluation of these two simpler contributions is detailed in the next section. The nature of the hydrodynamic torque arising from the perturbed stratification depends crucially on $Pe$, and this more complicated calculation is given in sections \ref{sec:smallPe} and \ref{sec:largePe}, respectively.

\section{The spheroidal angular velocity due to the inertial and hydrostatic torque contributions} \label{sec:inertiastatic:torque}

The $O(Re)$ inertial torque in (\ref{recithm_Angvelfinal}) has recently been calculated for spheroids, both prolate and oblate, of an arbitrary aspect ratio\,(see \cite{Marath_JFM2015}). Although the analysis in \cite{Marath_JFM2015} pertains to the limit $Re \ll 1$, the results have been shown to remain qualitatively valid even for $Re$'s of order unity\,(see \cite{Mehlig_2020}). As mentioned above, this torque has a regular character, and the regularity may be seen from the convergence of the inertial volume integral in (\ref{recithm_Angvel}) based on a leading order Stokesian estimate for the integrand. As argued in \cite{Marath_JFM2015}, the inertial acceleration ${\boldsymbol u}^{(1)} \cdot \nabla {\boldsymbol u}^{(1)} \sim \hat{\boldsymbol U} \cdot \nabla {\boldsymbol u}^{(1)} \sim O(1/r^2)$ for distances large compared to $L$, or $r \gg 1$ in dimensionless terms, on using ${\boldsymbol u}^{(1)} \sim O(1/r)$ for the Stokeslet field due to the translating spheroid. The test velocity field ${\boldsymbol u}^{(2)}$ due to the rotating spheroid has the character of a rotlet (and stresslet) in the far-field, and is therefore $O(1/r^2)$. This leads to an integrand that decays as $\hat{\boldsymbol U} \cdot \nabla {\boldsymbol u}^{(1)} \cdot {\boldsymbol u}^{(2)} \sim O(1/r^4)$ for $r \gg 1$, implying a convergent volume integral. This volume integral has been evaluated in closed form using spheroidal coordinates in \cite{Marath_JFM2015}. For the prolate case, the spheroidal coordinates $(\xi,\eta,\phi)$ are defined by the relations: $x_1 + \mathrm{i} x_2 = d\bar{\xi}\bar{\eta} \exp(i \phi)$, $x_3 = d\xi\eta$, with the $3$-axis of the Cartesian system aligned with the spheroid axis of symmetry. Here, $1 \leq \xi < \infty$, $|\eta| \leq 1$ and $0 \leq \phi < 2\pi$, with $\bar{\xi} = (\xi^2-1)^{\frac{1}{2}}$ and $\bar{\eta} = (1-\eta^2)^{\frac{1}{2}}$. The constant-$\xi$ surfaces correspond to confocal prolate spheroids and the constant-$\eta$ surfaces to confocal two-sheeted hyperboloids, both with the interfoci distance $2d$, and the constant-$\phi$ surfaces are planes passing through the axis of symmetry. The corresponding expressions for the oblate case may be obtained by the substitutions $d \leftrightarrow -\mathrm{i}d$, $\xi \leftrightarrow \mathrm{i}\bar{\xi}$. In either case, the spheroid is the surface $\xi = \xi_0$, its aspect ratio being given by $\kappa = \frac{\xi_0}{\bar{\xi_0}}$ and  $\frac{\bar{\xi}_0}{\xi_0}$ for the prolate and oblate cases; thus, the near-spherical limit\,($\kappa \rightarrow 1$) for either prolate or oblate spheroids corresponds to $\xi_0 \rightarrow \infty$, while the slender fiber\,($\kappa \rightarrow \infty$) and flat disk\,($\kappa \rightarrow 0$) limits correspond to $\xi_0 \rightarrow 1$. The fluid domain in the volume integrals in (\ref{recithm_Angvelfinal}) corresponding to $\xi \geq \xi_0$. 

\begin{figure}
\centering
\includegraphics[width=0.8\linewidth]{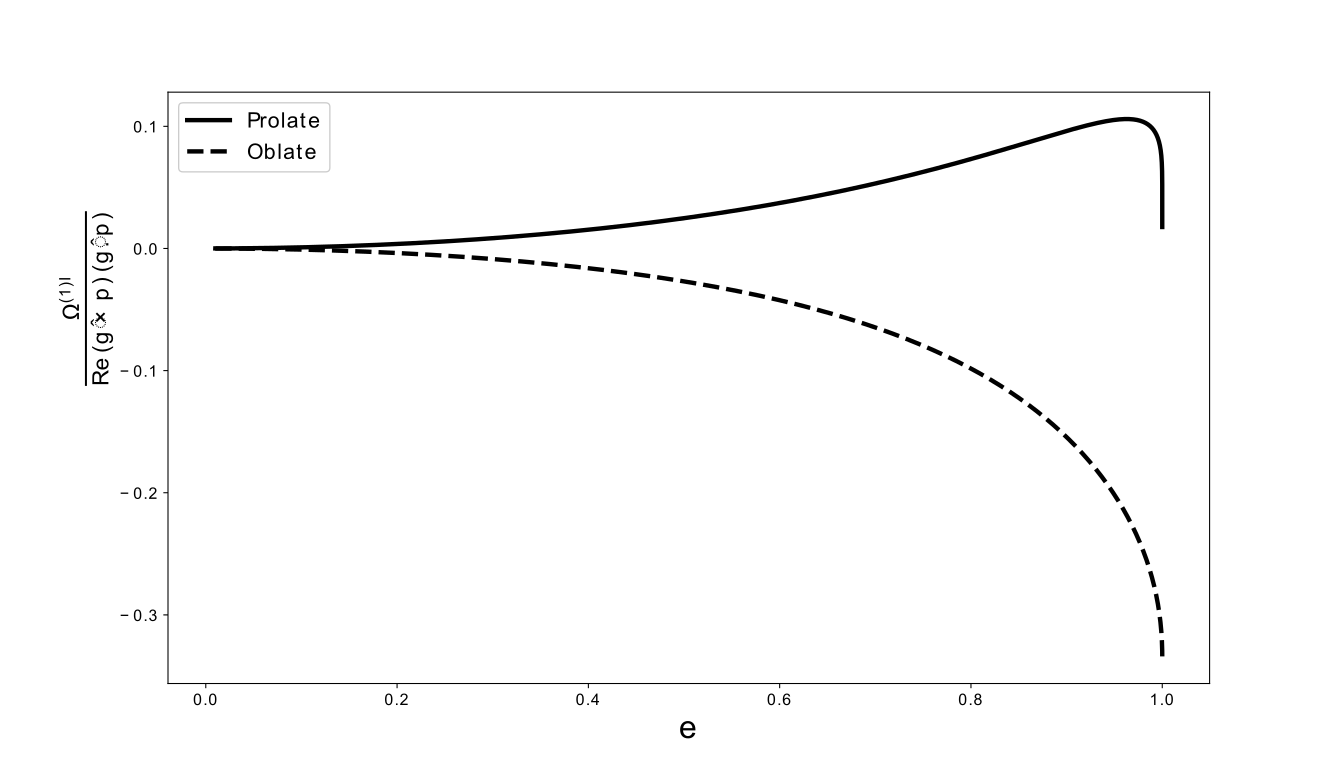}
\caption{The functions $\frac{F^p_I(\xi_0)X_A}{Y_C Y_A}$ and $\frac{F^o_I(\xi_0)X_A}{Y_C Y_A}$, that characterize the aspect-ratio-dependence of the inertial contributions to the angular velocities of prolate and oblate spheroids, plotted as a function of the spheroid eccentricity.}\label{fig:inertia}
\end{figure}

For a prolate spheroid, the actual velocity field ${\boldsymbol u}^{(1)}$ and the test velocity field tensor ${\boldsymbol U}^{(2)}$ in (\ref{recithm_Angvelfinal}), may be written in terms of the appropriate decaying vector spheroidal harmonics\,(see \cite{Marath_JFM2015};\cite{Marath_JFM2016}) as: 
\begin{align}
 {\boldsymbol u}^{(1)} = -\left(\frac{\hat{\boldsymbol{U}}.\boldsymbol{1}_3}{\xi_0 Q_1^1(\xi_0)+Q_0^0(\xi_0)} \right)\boldsymbol{S}_{1,0}^{(3)} -\left(\frac{\hat{\boldsymbol{U}}.\boldsymbol{1}_1}{3 Q_0^0(\xi_0)- \xi_0 Q_1 ^0(\xi_0)} \right)(\boldsymbol{S}_{1,1}^{(3)}-\boldsymbol{S}_{1,-1}^{(3)}), \label{StokesTranslation_vel}
\end{align}
\begin{align}
{\boldsymbol U}^{(2)}=\boldsymbol{1}_2\left(\frac{d(2\xi_0^2-1)(\boldsymbol{S}_{1,1}^{(2)}-\boldsymbol{S}_{1,-1}^{(2)})}{\left[2 \xi_0 Q_1^0(\xi_0)-\sqrt{\xi_0^2-1}Q_1^1(\xi_0)\right]}+\frac{d\left[\xi_0 Q_1^1(\xi_0)+2 \sqrt{\xi_0^2-1}Q_1^0(\xi_0)\right](\boldsymbol{S}_{2,1}^{(3)}-\boldsymbol{S}_{2,-1}^{(3)})}{Q_2^1(\xi_0)\left[2 \xi_0 Q_1^0(\xi_0)-\sqrt{\xi_0^2-1}Q_1^1(\xi_0)\right]}\right)  \label{TestVel_tensor}.
\end{align}
The $\boldsymbol{S}^{(3)}_{t,s}$'s and $\boldsymbol{S}^{(2)}_{t,s}$'s in (\ref{StokesTranslation_vel}) and (\ref{TestVel_tensor}) denote the decaying\,(biharmonic and harmonic) vectorial solutions of the Stokes equations in spheroidal coordinates, and are given by the following expressions:
\begin{align}
\boldsymbol{S}_{1,0}^{(3)} =&\left[ {(x_1\boldsymbol{1}_1+x_2\boldsymbol{1}_2+x_3\boldsymbol{1}_3)}\frac{\partial}{\partial x_3} Q_0^0(\xi) -\left( Q_0^0(\xi) + d\xi_0^2\left(\frac{\partial}{\partial x_3} \left[Q^0_1(\xi)P^0_1(\eta)\right]\right) \right) {\boldsymbol 1}_3 \nonumber \right. \\ & \left. - d(\xi_0^2-1)\left({\boldsymbol 1}_1 \frac{\partial }{\partial x_1}\left[Q^0_1(\xi)P^0_1(\eta)\right] + {\boldsymbol 1}_2 \frac{\partial }{\partial x_2}\left[Q^0_1(\xi)P^0_1(\eta)\right]\right) \right], \label{S3_10} \\
\boldsymbol{S}_{1,1}^{(3)} -\boldsymbol{S}_{1,-1}^{(3)}=&\left[ \!-2\!\left(\!\!{(x_1\boldsymbol{1}_1+x_2\boldsymbol{1}_2+x_3\boldsymbol{1}_3})\frac{\partial Q_0^0(\xi)}{\partial x_1}\! -\! Q_0^0(\xi) {\boldsymbol 1}_1\!\! \right)- {\boldsymbol 1}_3 d\xi_0^2 \frac{\partial}{\partial x_3}(P_1^1(\eta)Q_1^1(\xi)\cos \phi)\! \right. \nonumber \\ &\left. -\! d (\xi_0^2-1)\left(\!\!
{\boldsymbol 1}_1\frac{\partial}{\partial x_1}\! +\! {\boldsymbol 1}_2\frac{\partial }{\partial x_2}\!\right)\!(P_1^1(\eta)Q_1^1(\xi)\cos \phi) \right], \label{S3_111m1} \\
\boldsymbol{S}_{1,1}^{(2)}+\boldsymbol{S}_{1,-1}^{(2)} =&\left[2 P_1^0(\eta) Q_1^0(\xi) \mathbf{1}_1+P_1^1(\eta)Q_1^1(\xi) \cos(\phi)\mathbf{1}_3\right], \label{S2_111m1}\\
\boldsymbol{S}_{2,1}^{(3)}-\boldsymbol{S}_{2,-1}^{(3)} =&\left[(x_1\boldsymbol{1}_1+x_2\boldsymbol{1}_2+x_3\boldsymbol{1}_3)\frac{\partial}{\partial x_3}(P_1^1(\eta) Q_1^1(\xi) \sin (\phi)) -\frac{d\xi_0^2}{3}\nonumber \right. \\&\left.\frac{\partial}{\partial x_3}(P_2^1(\eta)Q_2^1(\xi)\cos(\phi))\mathbf{1}_{3}-\frac{d(\xi_0^2-1)}{3}\left(\mathbf{1}_{1} \frac{\partial}{\partial x_1}+\mathbf{1}_{2} \frac{\partial}{\partial x_2} \right)(P_2^1(\eta) Q_2^1(\xi)\cos (\phi)) \right], \label{S3_212m1}
\end{align}
with the $P_t^s(\eta)$ and $Q_t^s(\xi)$ being the associated Legendre functions of the first and second kind, respectively\,(for $s = 0$, the $P_t^s(\eta)$ correspond to the usual Legendre polynomials). The resulting volume integration may then be carried out analytically, and the inertial angular velocity ($\boldsymbol{\Omega}^{(1)I}$) is given by:
\begin{align}
\Omega_i^{(1)I}=& Re \left[ \frac{F^p_I(\xi_0)X_A}{Y_C Y_A} (\, \epsilon_{ijk}\hat{g}_j p_k \, \hat{g}_lp_l)\right],\label{eqn:prolate_inertialtorque}
\end{align}
for prolate spheroids. The corresponding expression for the oblate case may be obtained by the aforementioned substitutions viz. $d \leftrightarrow -\mathrm{i}d$, $\xi_0 \leftrightarrow \mathrm{i}\bar{\xi}_0\,$in the dimensional angular velocity, and is given by:
\begin{align}
\Omega_i^{(1)I}=& Re\left[ \frac{F^o_I(\xi_0)X_A}{Y_C Y_A}(\epsilon_{ijk}\hat{g}_j p_k \hat{g}_lp_l) \right]. \label{eqn:oblate_inertialtorque}
\end{align}
The expressions for $F^p_I(\xi_0)$ and $F^o_I(\xi_0)$, as functions of the spheroid eccentricity\,($e=1/\xi_0$), were first obtained by \cite{Marath_JFM2015}, and are given in Appendix \ref{appendixresistance}. The inertial angular velocity given by (\ref{eqn:prolate_inertialtorque}) and  (\ref{eqn:oblate_inertialtorque}) orients sedimenting spheroids broadside-on regardless of $\kappa$. The combination of the aspect-ratio-dependent functions, $F^{p/o}_I(\xi_0) X_A/(Y_CY_A)$, that multiplies $Re(\hat{\boldsymbol g} \cdot{\boldsymbol p})(\hat{\boldsymbol g} \wedge {\boldsymbol p})$, and that determines the $\kappa$-dependence of the inertial angular velocities above, are plotted as a function of the eccentricity in Figure \ref{fig:inertia}, for both the prolate and oblate cases. One obtains the expected $O(1/\xi_0^2)$ scaling in the near-sphere limit ($\xi_0 \rightarrow \infty$); at the other extreme($\xi_0 \rightarrow 1$), the inertial angular velocity approaches zero as $O[\ln (\xi_0-1)]^{-1}$ in the slender fiber limit, consistent with viscous slender body theory\,(\cite{KhayatCox_1989},\cite{SubKoch_2005}), while remaining finite in the limit of a flat disk.

\begin{figure}
\centering
\includegraphics[width=0.8\linewidth]{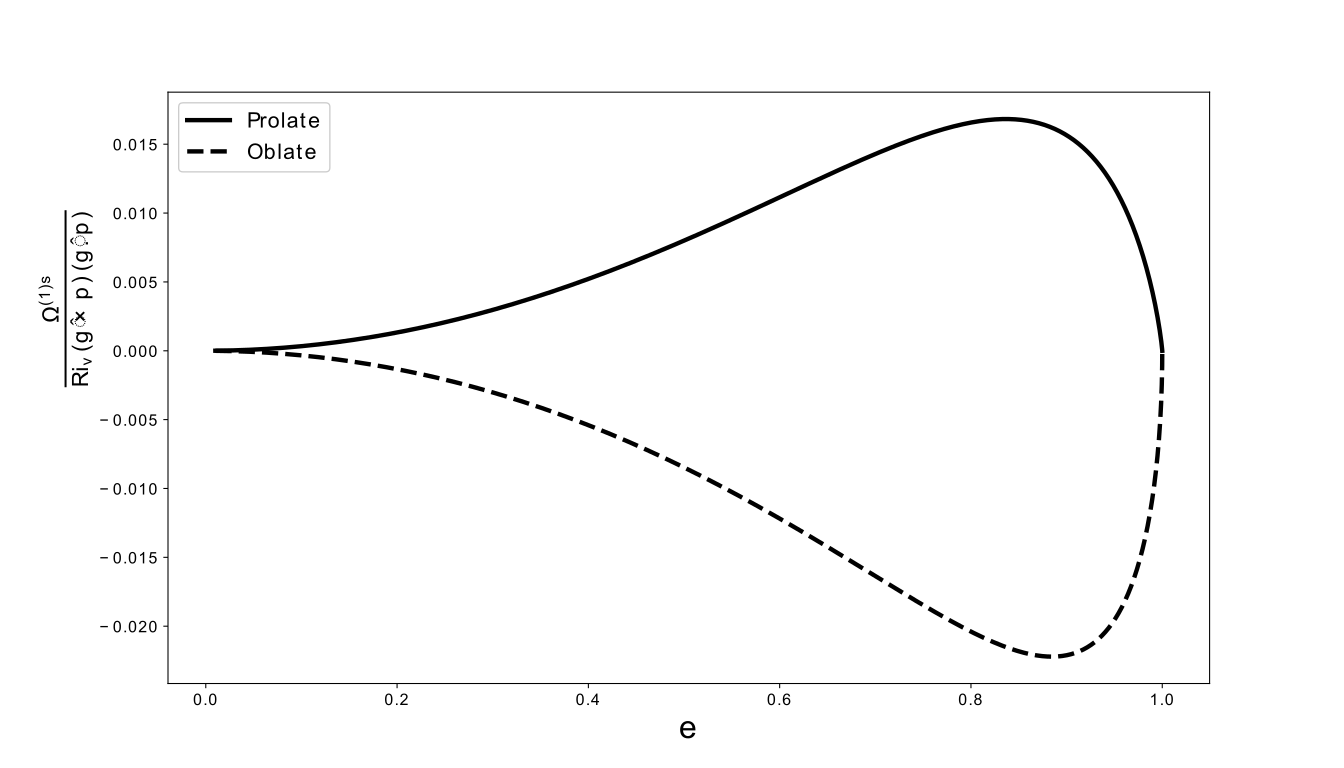}
\caption{The aspect ratio dependence of the hydrostatic contributions to the angular velocities of prolate and oblate spheroids.}\label{fig:hydrostatic}
\end{figure}

The hydrostatic component of the stratification torque is also readily evaluated in spheroidal coordinates. For the prolate case, the dimensionless position vector that appears in (\ref{hydrostat:torque}) is given by $\boldsymbol{x} = \frac{\bar{\xi}_0}{\xi_0}\bar{\eta}(\cos \phi {\boldsymbol 1}_1 + \sin \phi {\boldsymbol 1}_2) + \eta {\boldsymbol 1}_3$, and the unit normal is $\boldsymbol{n} = {\boldsymbol 1}_\xi = \frac{\xi_0\bar{\eta}}{\sqrt{\xi_0^2 -\eta^2}}(\cos \phi {\boldsymbol 1}_1 + \sin \phi {\boldsymbol 1}_2) + \frac{\bar{\xi_0}\eta}{\sqrt{\xi_0^2 -\eta^2}} {\boldsymbol 1}_3$. Using these expressions, and the areal element $dS = h_\eta h_\phi d\eta d\phi$, with $h_\eta= \frac{\sqrt{\xi_0^2- \eta^2}}{\xi_0\bar{\eta}}$ and $h_\phi = \frac{\bar{\xi}_0\bar{\eta}}{\xi_0}$, one obtains the angular velocity due to the hydrostatic torque as
\begin{align}
\Omega_i^{(1)s} = &\,Ri_v\frac{4\pi}{15Y_C}\frac{1-\xi_0^2 }{\xi_0^4}(\epsilon_{ijk} \hat{g}_j p_k)(\hat{g}_lp_l), \label{eqn:prolate_hydrostattorque}
\end{align}
for the prolate case, and using the transformations mentioned above,
\begin{align}
\Omega_i^{(1)s} =&\,Ri_v\frac{4\pi}{15Y_C}\frac{\sqrt{\xi_0^2 -1}}{\xi_0^3}(\epsilon_{ijk} \hat{g}_j p_k)(\hat{g}_lp_l), \label{eqn:oblate_hydrostattorque}
\end{align}
for the oblate case. The angular velocities given by (\ref{eqn:prolate_hydrostattorque}) and (\ref{eqn:oblate_hydrostattorque}) also orient the spheroid broadside-on like the inertial torque above. The aspect-ratio-dependent functions that multiply $Ri_v(\hat{\boldsymbol g} \cdot {\boldsymbol p})(\hat{\boldsymbol g} \wedge {\boldsymbol p})$ in (\ref{eqn:prolate_hydrostattorque})and (\ref{eqn:oblate_hydrostattorque}) are plotted as functions of $\xi_0$ in Fig. \ref{fig:hydrostatic}. Since the hydrostatic torque is only a function of the particle geometry, these aspect ratio functions are algebraically small in both the near-sphere, and the slender fiber and flat-disk limits.

The inertial and hydrostatic angular velocities above have an identical angular dependence, of the form $(\hat{\boldsymbol g} \cdot {\boldsymbol p})(\hat{\boldsymbol g} \wedge {\boldsymbol p})$, one which is easily inferred based on the requirement that the angular velocity be a pseudovector quadratic in $\hat{\boldsymbol g}$\,(\cite{Marath_JFM2015}). The dependence implies that the maximum angular velocity occurs midway between the horizontal\,($\hat{\boldsymbol g} \cdot {\boldsymbol p} = 0$)  and vertical \,($\hat{\boldsymbol g} \cdot {\boldsymbol p} = 1$) orientations. The hydrostatic torque arises because the point of action of the upward buoyant force, the center of mass of the equivalent stratified fluid blob\,(in the Archimedean interpretation), lies below the geometric center through which the weight of the spheroid acts downward. The two forces therefore constitute a couple that turns the spheroid broadside-on. The broadside-on nature of the inertial torque is on account of `wake-shielding' - the wake associated with the front portion of the spheroid shields the rear, which catches up with the front as a result. As pointed out in \cite{Marath_JFM2015}, this not literally true for small $Re$.  A signature of the wake arises only on length scales greater than $O(LRe^{-1})$, in contrast to the scaling arguments above which show that the $O(Re)$ inertial torque arises from fluid inertial forces in a region of $O(L^3)$\,(the inner region) around the sedimenting spheroid. Nevertheless, the velocity field in the inner region reflects the asymmetry of the outer Oseen field, and the sense of rotation remains the same for small $Re$. Importantly, the broadside-on nature of the inertial and hydrostatic torques imply that the transition from broadside-on to edgewise settling, observed in the recent experiments\,(see \cite{mercier_2020}) discussed in the introduction, must depend entirely on the hydrodynamic component of the stratification torque, that is, the second term within square brackets on the right hand side in (\ref{recithm_Angvelfinal}). While the calculation above shows the hydrostatic component to be $O(Ri_v)$, consistent with the nominal order in (\ref{recithm_Angvelfinal}), this is not true of the hydrodynamic component. As will be shown in section \ref{hydrodyn:torque} below, the hydrodynamic component scales as $O(Ri_v)$ only for $Pe \ll 1$ when the dominant contribution to the associated torque integral comes from length scales of $O(L)$ similar to the inertial torque above. In the opposite limit of $Pe \gg 1$, and for the so-called Stokes stratification regime corresponding to $Re \ll Ri_v^{\frac{1}{3}}$\,(see \cite{Arun_2020}), the dominant contributions to the torque integral arise from much larger length scales of $O(LRi_v^{-\frac{1}{3}})$, and the hydrodynamic component scales as $O(Ri_v^{\frac{2}{3}})$, being much larger than the hydrostatic component above.

Before proceeding with the calculation of the hydrodynamic component of the stratification torque, it is worth remarking on the nature of the coupling between the inertial and stratification torque contributions that is not obvious from the formal result (\ref{recithm_Angvel}) above, where they appear as separate additive contributions. On account of the convergent volume integral, the $O(Re)$ inertial angular velocity, as given by (\ref{eqn:prolate_inertialtorque}) and (\ref{eqn:oblate_inertialtorque}), only involves the Stokesian fields in a homogeneous ambient, and is evidently independent of the ambient stratification. The correction to this leading order estimate is dependent on the nature of the ambient stratification, however, even within the Boussinesq framework. To see this, we return to the inertial volume integral, and estimate the next correction. Recall that the angular velocities in (\ref{eqn:prolate_inertialtorque}) and (\ref{eqn:oblate_inertialtorque}) were based on the approximating the volume integral by Stokesian estimates, and the torque contribution at the next order requires one to examine the next term in the small-$Re$ expansion for the velocity field in problem 1. Writing ${\boldsymbol u}^{(1)} = {\boldsymbol u}^{(10)}+ Re\,{\boldsymbol u}^{(11)}$, ${\boldsymbol u}^{(10)}$ is the Stokesian approximation given by (\ref{StokesTranslation_vel}) above and is $O(1/r)$ for $r \gg 1$, while ${\boldsymbol u}^{(11)}$ remains $O(1)$ in the far-field. The latter, of course, implies that the above regular expansion breaks down at length scales of $O(LRe^{-1})$, a manifestation of the singular nature of inertia in an unbounded domain\,(the so-called Whitehead's paradox; see \cite{Leal_book}). Provided one assumes buoyancy forces to dominate the inertial ones on scales much smaller than the inertial screening length\,(of $O(LRe^{-1}))$, the above far-field estimate of ${\boldsymbol u}^{(11)}$ may still be used to estimate the correction to the leading $O(Re)$ contribution. The $O(Re^2)$ inertial acceleration is now $\hat{\boldsymbol U} \cdot \nabla {\boldsymbol u}^{(11)} \sim O(1/r)$, and using ${\boldsymbol u}^{(2)} \sim O(1/r^2)$ , the resulting volume integral, at $O(Re^2)$, is logarithmically divergent. The divergence will be cut off at the stratification screening length that is $O(Ri_vPe)^{-\frac{1}{4}}$ for $Pe \ll 1$\,(\cite{ardekani_2010},\cite{list_laminar_1971}), and $O(Ri_v^{-\frac{1}{3}})$ for $Pe \gg 1$\,(\cite{Arun_2020}), implying that the next correction to the inertial angular velocity is $O[Re^2\ln (Ri_v Pe)^{-\frac{1}{4}}]$ for $Pe \ll 1$ and $O[Re^2\ln Ri_v^{-\frac{1}{3}}]$ for $Pe \gg 1$, and is thereby a function of the ambient stratification. For self consistency, one requires that both of the aforementioned stratification screening lengths be less than $O(Re^{-1})$, which translates to the requirement $Ri_v \gg Pe^{-1} Re^4$ for small $Pe$, and for one to be in the aforementioned Stokes stratification regime\,($Ri_v^{\frac{1}{3}} \gg Re$) for large $Pe$.

\section{The spheroidal angular velocity due to the hydrodynamic component of the stratification contribution} \label{hydrodyn:torque}

Owing to the differing character of the hydrodynamic component in the limits of small and large $Pe$, the calculations in these two asymptotic regimes are carried out in separate subsections below. Keeping in mind that $Pe = RePr$, $Pr$ being the Prandtl number, the small $Pe$ case doesn't necessarily place a restriction on $Pr$ which may either be small or large\,(although, large $Pr$ imposes a greater restriction on the smallness of $Re$ since $Re$ must now be smaller than $O(Pe^{-1})$). However, the limit of small $Re$ that characterizes the analysis here implies that the large $Pe$ case necessarily requires a large $Pr$ which may be realized in experiments that use salt as a stratifying agent.

\subsection{The hydrodynamic stratification torque in the diffusion-dominant limit\,($Pe \ll 1$)} \label{sec:smallPe}

In the limit $Pe \ll 1$, one may neglect the convective terms in the advection diffusion equation (\ref{phys:densitycons}), and the density perturbation $\rho'^{(1)}$ in the stratification torque integral therefore arises as a diffusive response to the no-flux condition that must be satisfied on the spheroid surface. As a result, the spheroid acts as a concentration-dipole singularity in the far-field\,($r \gg 1$), implying that $\rho'^{(1)}$ must decay as $O(1/r^2)$. Since the test velocity field ${\boldsymbol u}^{(2)}$ corresponding to the rotating spheroid also decays as $O(1/r^2)$, the integrand is $O(1/r^4)$ for $r \gg 1$. This decay is the same as that of the inertial integrand estimated above, and the integral for the hydrodynamic stratification torque is therefore convergent for small $Pe$, based on the leading order diffusive estimates above. Thus, the effects of stratification arise as a regular perturbation for small $Pe$, or said differently, the dominant contribution to the hydrodynamic component of the stratification torque  arises from buoyancy forces in a volume of $O(L^3)$ around the sedimenting spheroid. As a consequence and as is shown below, for $Pe \ll 1$, the hydrodynamic component is $O(Ri_v)$ similar to the hydrostatic component given in (\ref{eqn:prolate_hydrostattorque}) and (\ref{eqn:oblate_hydrostattorque}) above. Note that the $O(Ri_v)$ scaling implies that the associated (dimensional) angular velocity is independent of $U$. While this must be the case for the hydrostatic contribution, it turns out to be the case for the hydrodynamic component too, at small $Pe$, since the leading order density perturbation arises as a diffusive response, and is therefore independent of the ambient flow. Note that despite this $U$-independence, the torque does have a hydrodnamic character, in that it still arises from the flow induced by buoyancy forces. The apparent conflict arises only because $U$ is not the appropriate velocity scale for $Pe \ll 1$. Rather, using a density perturbation of $O(\gamma L)$, one obtains a buoyancy-driven velocity scale of $O(\gamma L^3 g/\mu)$ from the equations of motion, implying a spheroidal angular velocity of $O(\gamma L^2 g/\mu)$; this is $O(Ri_v)$ in units of $U/L$, the latter scale being used in the reciprocal theorem formulation in section \ref{sec:recithm}. A genuine dependence of the hydrodynamic torque contribution on $U$ arises at the next order, however, and an estimate of this contribution is obtained at the end of this subsection.

To determine the detailed dependence of the $O(Ri_v)$ hydrodynamic component on $\kappa$, one needs to solve for the density perturbation $\rho'^{(1)}$ which satisfies:
\begin{align}
\nabla^2 \rho'^{(1)} = 0.   \label{densitypert_Laplacian}
\end{align}
in the fluid domain $\xi \geq \xi_0$. The no-flux boundary condition on the spheroid surface\,($\xi = \xi_0$) may be written as ${\boldsymbol 1}_\xi \cdot {\boldsymbol \nabla} \rho'^{(1)} = -{\boldsymbol 1}_\xi \cdot \hat{\boldsymbol g}$ where we have used that ${\boldsymbol n} = {\boldsymbol 1}_\xi$, and the right hand side of the boundary conditon arises from the gradient of the linearly varying ambient density; there is the additional requirement of far-field decay viz. $\rho'^{(1)} \rightarrow 0$ for $\xi \rightarrow \infty$. Note that the linearity of the governing equation (\ref{densitypert_Laplacian}) and the boundary conditions in $\rho'^{(1)}$, and the linear dependence on $\hat{\boldsymbol g}$ of the surface boundary condition above, imply that $\rho'^{(1)}$ must be linear in $\hat{\boldsymbol g}$ at leading order for small $Pe$. From (\ref{recithm_Angvelfinal}), the hydrodynamic component of the stratification torque must therefore be quadratic in $\hat{\boldsymbol g}$. This implies that the hydrodynamic torque must have an angular dependence identical to the inertial and hydrostatic contributons, of the form $(\hat{\boldsymbol g} \wedge {\boldsymbol p})(\hat{\boldsymbol g} \cdot {\boldsymbol p}$), with a multiplicative pre-factor that is a function of $\kappa$. Thus, for small $Pe$, the relative magnitudes of the hydrostatic and hydrodynamic components of the stratification torque is independent of the spheroid orientation and $Ri_v$, and only a function of $\kappa$.

To solve (\ref{densitypert_Laplacian}), we write down the explicit form of the no-flux boundary condition, in prolate spheroidal coordinates:
\begin{align}
\frac{\partial \rho'^{(1)}}{\partial \xi}|_{\xi = \xi_0} =&\,\frac{(P_1^1(\eta) e^{\mathrm{i}\phi} - 2 P_1^{-1}(\eta)e^{\mathrm{i}\phi} )}{2\bar{\xi}_0} \sin \psi - \frac{P^0_1(\eta)}{\xi_0} \cos \psi,  \label{surf_BC:spheroidal}
\end{align}
where $\psi$ denotes the angle between ${\boldsymbol p}$ and ${\boldsymbol g}$. In (\ref{surf_BC:spheroidal}), $P_1^1(\eta) = \bar{\eta}$, $P_1^{-1}(\eta) =-\frac{\bar{\eta}}{2}$ and $P_1^0(\eta) = \eta$. The $\eta$-dependence of the solution is imposed by that of the boundary condition above. This, and the fact that the Laplacian is separable in spheroidal coordinates with the eigenfunctions in $\xi$ and $\eta$ being the associated Legendre functions of the second and the first kind, respectively, points to the following ansatz for $\rho'^{(1)}$:
\begin{align}
\rho'^{(1)} =&\, A_{1,1}Q_1^1(\xi)P_1^1(\eta) e^{\mathrm{i}\phi} + A_{1,-1} Q_1^{-1}(\xi)P_1^{-1}(\eta)e^{-\mathrm{i}\phi}  + A_{1,0}Q_1^0(\xi)P_1^0(\eta), \label{smallPe:rhoansatz}
\end{align}
where $Q_1^0(\xi) = \xi  \coth ^{-1}\xi -1$, $Q_1^1(\xi) =\left[\left(\xi ^2-1\right) \coth ^{-1}\xi -\xi \right]/{\bar{\xi}}$ and $Q_1^{-1}(\xi) =Q_1^1(\xi)/2$, and the $A_{i,j}$'s are the unknown constants. Substitution of (\ref{smallPe:rhoansatz}) leads to the following expressions for the $A_{i,j}$'s:
\begin{align}
A_{1,1} =&\, \frac{\bar{\xi}_0^2}{4-2\xi_0^2 + 2\xi_0\bar{\xi}_0^2\coth^{-1}\xi_0}\sin \psi,  \label{eq:A11} \\
A_{1,-1} =&\,-4A_{1,1}, \label{eq:A1m1} \\
A_{1,0} =&\,\frac{\bar{\xi}_0^2}{\xi_0(\xi_0 - \bar{\xi}_0^2\coth^{-1}\xi_0)}\cos \psi. \label{eq:A10}
\end{align}
The first two terms in (\ref{smallPe:rhoansatz}) correspond to the density perturbation induced by a spheroid oriented transversely to gravity, while the third term corrresponds to that induced by a spheroid aligned with gravity. The expressions for the $A_{i,j}$'s above are therefore consistent with the underlying linearity of the problem, in that the perturbation induced by an arbitrarily oriented spheroid is obtained as a superposition of the transverse and longitudinal problems, the factors involved in this superposition being $\sin \psi$ and $\cos \psi$, respectively.

\begin{figure}
\centering
\includegraphics[width=0.8\linewidth]{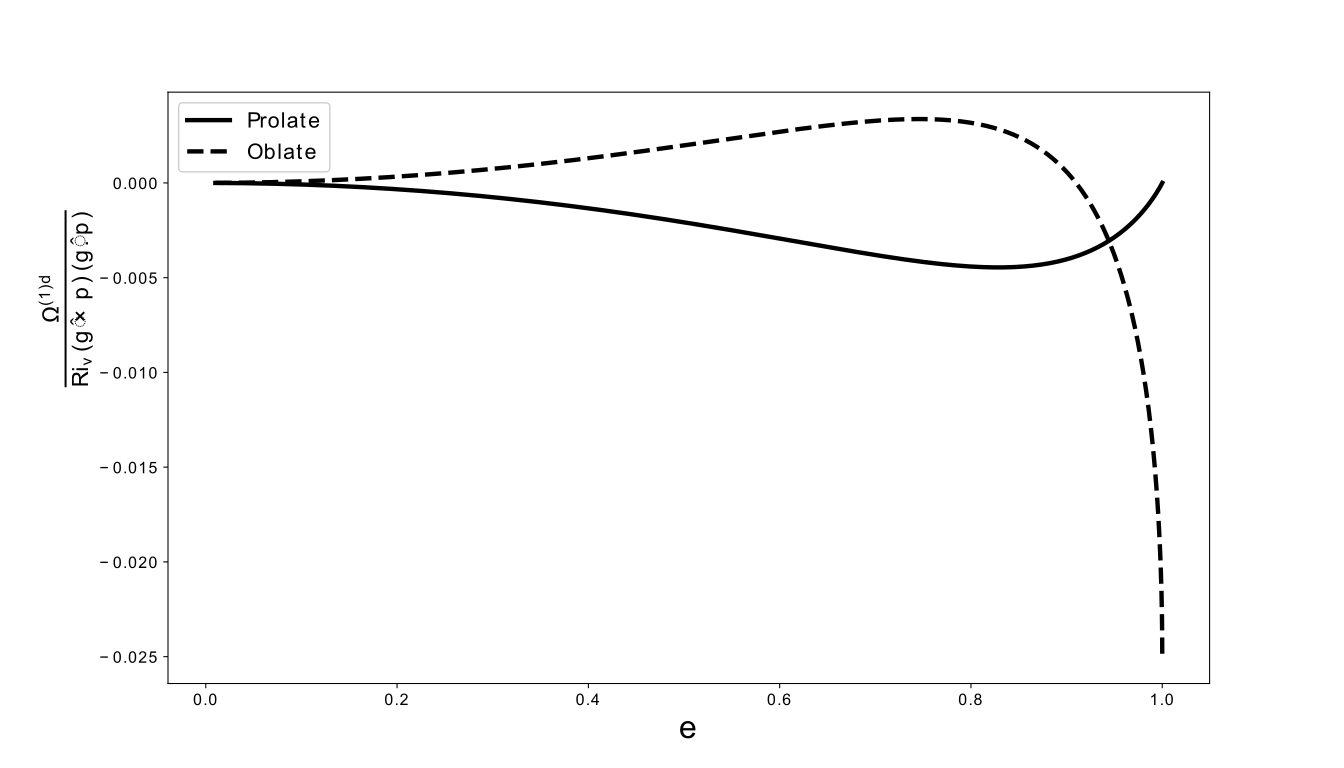}
\caption{The aspect ratio dependence of the angular velocities of prolate and oblate spheroids due to the hydrodynamic torque in the diffusion-dominant limit.}\label{fig:hydrodynamic}
\end{figure}

Use of (\ref{smallPe:rhoansatz}), together with (\ref{eq:A11}-\ref{eq:A10}), in the integral for the hydrodynamic stratification torque in (\ref{recithm_Angvelfinal}), leads to the following expression for the spheroid angular velocity:
\begin{align}
\Omega_i^{(1)d} =&Ri_v  \frac{F^p_s(\xi_0)}{Y_C}\,\, (\epsilon_{ijk} \hat{g}_j p_k)\,(\hat{g}_lp_l)\,  \label{prolate_dynangvel},
\end{align}
with 
\begin{align}
\scriptstyle F^p_s(\xi_0)=&\,\scriptstyle{ {\frac{2\pi  {\bar{\xi_0}}^2  \left(\left(7 \xi_0^5-7 \xi_0^3-2\xi_0\right)+\bar{\xi_0}^2 \coth ^{-1}\xi_0 \left(-12 \xi_0^4+6 \xi_0^2+\xi_0 \coth ^{-1}\xi_0 \left(3 \xi_0^4+2 \left(\xi_0^5-\xi_0 \right) \coth ^{-1}\xi_0-6 \xi_0^2+7\right)-2\right)\right)}{15 \xi_0^5 \left(\xi_0-\bar{\xi_0}^2 \coth ^{-1}\xi_0\right) \left(-\xi_0^2+\bar{\xi_0}^2 \xi_0 \coth ^{-1}\xi_0+2\right) \left(\left(\xi_0^2+1\right) \coth ^{-1}\xi_0-\xi_0\right)}}}, \label{smallPe:prolateaspectratio}
\end{align}
for prolate spheroids, and via the transformation mentioned in section \ref{sec:inertiastatic:torque}, to
\begin{align}
 \scriptstyle F^o_s(\xi_0)=&\, \scriptstyle{\frac{2\pi   \left(\left(7 \xi_0^4-7 \xi_0^2-2\right) \bar{\xi}_0+2 \left(\xi_0^4-3 \xi_0^2+2\right) \xi_0^4 \left(\cot ^{-1}\bar{\xi}_0\right)^3-2 \left(6 \xi_0^4-9 \xi_0^2+4\right) \xi_0^2 \cot ^{-1}\bar{\xi}_0+\left(3 \xi_0^4+4\right) \bar{\xi}_0 \xi_0^2 \left(\cot ^{-1}\bar{\xi}_0\right)^2\right)}{15 \xi_0^3 \left(\xi_0^2 \cot ^{-1}\left(\bar{\xi}_0\right)-\bar{\xi}_0\right) \left(\left(\xi_0^2-2\right) \left(\cot ^{-1}\bar{\xi}_0\right)-\bar{\xi}_0\right) \left(-\xi_0^2+\bar{\xi}_0 \xi_0^2 \cot ^{-1}\left(\bar{\xi}_0\right)-1\right)}},   \label{smallPe:oblateaspectratio}
\end{align}
for the oblate case. Fig. \ref{fig:hydrodynamic} shows a plot of the aspect-ratio-dependent functions given by (\ref{smallPe:prolateaspectratio}) and (\ref{smallPe:oblateaspectratio}). The hydrodynamic component of the stratification torque given by (\ref{smallPe:prolateaspectratio} always orients a prolate spheroid edgewise for small $Pe$. Interestingly, Fig. \ref{fig:hydrodynamic} shows that (\ref{smallPe:oblateaspectratio}) changes sign below a critical aspect ratio $\kappa_c \approx 0.17\,(\xi_0 \approx 0.9)$, and therefore, the hydrodynamic component acts to orient oblate spheroids, with aspect ratios lower than the aforementioned threshold, broadside on.

\begin{figure}
\centering
\includegraphics[width=\linewidth]{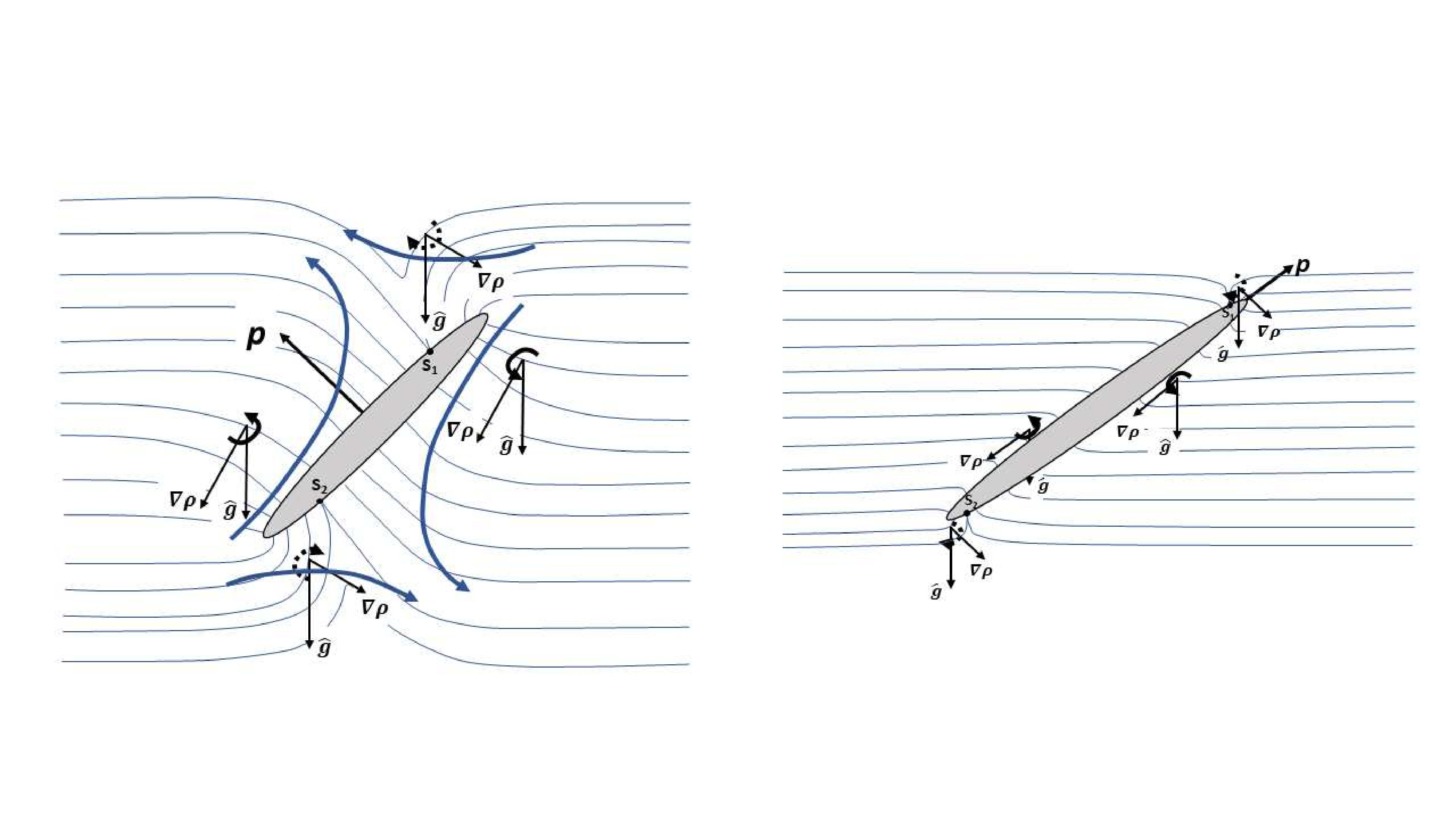}
\caption{Figures (a) and (b) depict the baroclinically-driven flow, for small $Pe$, that is responsible for the rotation of an (a) oblate, and a (b) prolate, spheroid in a stably stratified ambient. The curved arrows denote the sense of the baroclinically induced vorticity in the different quadrants of the fluid domain, with vorticities corresponding to anticlockwise and clockwise senses of rotation being denoted by solid and dashed lines; the blue contours denote the deformed iso-pycnals around each of the spheroids.}
\label{smallPe:mechanism}
\end{figure}

The hydrodynamic stratification torque arises due to the flow associated with the baroclinic source of vorticity, although the reciprocal theorem formulation used here bypasses the explicit calulation of this flow. The vorticity arises from the tilting of the iso-pycnals to the vertical\,(the direction of gravity) due to the requirement that they meet the spheroidal surface in a normal orientation, consistent with a no-flux constraint. A sketch of the deformed iso-pycnals in the plane $\phi = 0,\pi$, and the resulting sign of the baroclinc vorticity field\,($\propto \nabla \rho \wedge \hat{\boldsymbol g}$) in different regions of the fluid domain, appears in Figure \ref{smallPe:mechanism}, for both prolate and oblate spheroids. The baroclinically induced Stokesian flow has a dipolar character, and the relative sizes of the different flow quadrants are set by the pair of singular iso-pycnals that meet the spheroid surface in the points $S_1$ and $S_2$. This pair separates the iso-pycnals that do not meet the spheroid surface from those that do. The baroclinically induced flow on account of diffusive iso-pycnal tilting has been known for a long time, having originally been proposed in the oceanic context where the induced flow has a boundary layer character on account of the dominance of inertia\,(\cite{Phillips_1970},\cite{Wunsch_1970}). Such a flow has also been examined in the Stokesian context more recently\,(\cite{AliasPage_2017}), although only for the case of a horizontal circular cylinder wherein symmetry precludes a torque contribution. That there must be a torque on an inclined spheroid, due to the aforementioned baroclinic flow, is obvious. While the sense of the torque\,(broadside-on vs edgewise) is not readily evident, one may nevertheless rationalize the scalings observed for the extreme aspect ratio cases. Figure \ref{fig:hydrodynamic} shows that the angular velocity remains finite in the limit of a flat disk\,($\kappa \rightarrow 0$) which is consistent with the iso-pycnals being perturbed in a volume of $O(L^3)$ around the spheroid in this limit, with the density perturbation being $O(\gamma L)$, and the test velocity field ${\boldsymbol U}^{(2)}$ being $O(1)$ in this region; the points $S_1$ and $S_2$ remain bounded away from the edges of the flat disk. On the other hand, the angular velocity approaches zero as $O(\xi_0-1)$ in the limit of a slender fiber, with the points $S_1$ and $S_2$ now moving towards the ends of the fiber. The dominant contribution continues to come from a volume of $O(L^3)$. But, while ${\boldsymbol U}^{(2)}$ is $O[\ln(\xi_0-1)]^{-1}$, the density perturbation in this region is algebraically small. The slender fiber only perturbs the iso-pycnals in a thin $O(d^2L)$ shell around itself, with the density perturbation being $O(\gamma d)$ in this region. Further, each cross-section of the fiber acts as a 2D concentration dipole, implying that the density perturbation decays as $O(1/r)$ for $r$ much greater than $d$, and is therefore $O(\gamma d)(d/L)$ for $r \sim O(L)$. Using these estimates, and dividing by the $O[\ln(\xi_0-1)]^{-1}$ rotational resistance for a slender fiber leads to the aforementioned scaling for the fiber rotation due to the hydrodynamic contribution of the stratification torque.

We now examine the assumption of a quasi-steady response for the disturbance fields, implicit in the derivation of the inertial and low-$Pe$ stratification torques above, in order to ensure self-consistency. The time scales for setting up the steady disturbance velocity and density fields, on length scales of $O(L)$\,(the region that contributes dominantly to the leading $O(Ri_v)$ torque), are $O(L^2/\nu)$ and $O(L^2/D)$, respectively; the ratios of these time scales to the time scale of spheroid rotation come out to be $O(Re^2)$ and $O(ReRi_v)$ for the velocity field, and $O(PeRe)$ and $O(PeRi_v)$ for the density field, depending on whether $Re \gg Ri_v$ or vice versa. All of these ratios are asymptotically small in the limit of small $Pe$, making the quasi-steady state assumption a self-consistent one. Thus, for $Pe \ll 1$, all three contributions to the spheroid angular velocity that appear in (\ref{recithm_Angvelfinal}) have a regular character, and therefore, the same dependence on the spheroid orientation viz. $\sin \psi \cos \psi$ with $\psi$ being the angle between ${\boldsymbol p}$ and ${\boldsymbol g}$ as defined above. The inertial contribution is $O(Re)$, while both hydrodynamic and hydrostatic components of the stratification contribution are $O(Ri_v)$, with the hydrodynamic component alone acting to orient the spheroid edgewise for prolate spheroids of arbitrary $\kappa$ and oblate spheroids with $\kappa > 0.17 $. Therefore, oblate spheroids with $\kappa < 0.17$ will certainly orient broadside-on for $Pe \ll 1$. Further, it is seen from Fig.  \ref{fig:hydrostatic} and Fig. \ref{fig:hydrodynamic} that the hydrodynamic component remains smaller in magnitude than the hydrostatic one in the edgewise-rotation regime, and therefore, a sedimenting spheroid, either prolate or oblate, is expected to settle in the broadside-on configuration, regardless of $\kappa$, for sufficiently small $Pe$.
 
As will be seen in section \ref{sec:largePe} below, the dominant length scale contributing to the hydrodynamic component of the stratification torque changes from $O(L)$ to $O(L Ri_v^{-\frac{1}{3}})$ with increasing $Pe$, and one expects the singular effect of convection to therefore already be evident for small but finite $Pe$. While the $O(Ri_v)$ hydrodynamic component above arises from the perturbation of the ambient stratification on length scales of $O(L)$, driven by the no-flux condition on the spheroid surface, an independent contribution arises from perturbation of the ambient stratification on much larger length scales due to weak convection effects\,(small $Pe$). To obtain an estimate for this latter torque contribution, we consider the correction to the leading order density perturbation, now denoted as $\rho^{(10)}$, in a manner similar to the velocity field examined in the earlier section; thus, one writes $\rho^{(1)} =  \rho^{(10)} + Pe\,\rho^{(11)}$, with $\nabla^2 \rho^{(11)} \sim Pe\,{\boldsymbol u}'^{(1)} \cdot \hat{\boldsymbol g}$. With ${\boldsymbol u}' \sim 1/r$ on length scales smaller than $O(LRe^{-1})$, one obtains $\rho^{(11)} \sim r$. The stratification torque at the next order is proportional to $Ri_v Pe\textstyle\int \rho^{(11)} \hat{\boldsymbol g} {\boldsymbol u^{(2)}}dV$, which turns out to diverge as $O(r^2)$, on using the above estimate for $\rho^{(11)}$. Cutting off the divergence at the small-$Pe$ stratification screening length of $O[L(Ri_vPe)^{-\frac{1}{4}}]$\,(\cite{ardekani_2010},\cite{list_laminar_1971}) would seem to lead to an $O(Ri_vPe)^{\frac{1}{2}}$ torque contribution. However, this contribution turns out to be identically zero on account of the fore-aft symmetry of the disturbance density field on scales of $[L(Ri_vPe)^{-\frac{1}{4}}]$\,(\cite{ardekani_2010},\cite{Arun_2020}). The fore-aft asymmetry of the density perturbation, necessary for a non-trivial torque contribution, requires inclusion of the $O(Pe)$ convective term\,(${\boldsymbol u} \cdot \nabla \rho'$) in (\ref{prob1:densityconsND}). It is well known that, for $Pe$ small but finite, this convective term becomes comparable to the diffusive term on length scales of $O(LPe^{-1})$, the mass/heat transfer analog of the inertial screening length\,(\cite{Leal_book}), and the torque scaling therefore depends on the relative magnitudes of the convective\,($LPe^{-1}$) and stratification\,($L(Ri_vPe)^{-\frac{1}{4}}$) screening lengths, which in turns depends on the relative magnitudes of $Pe$ and $Ri_v^{\frac{1}{3}}$; note that this criterion is analogous to the classification into Stokes and inertia stratification regimes based on the structure of the large-$Pe$ disturbance velocity field\,(\cite{Arun_2020}), except that $Pe$ now replaces $Re$. For $Pe  \ll Ri_v^{\frac{1}{3}}$, fore-aft asymmetric buoyancy forces acting on scales of order the stratification screening length lead to an $O(Ri_v^{\frac{1}{4}}Pe^{\frac{5}{4}})$ hydrodynamic torque contribution. In the opposite limit of $Pe \gg Ri_v^{\frac{1}{3}}$, one expects the dominant buoyancy forces to arise on length scales of $O(LPe^{-1})$. Interestingly, and in contrast to the aforesaid expectation, the dominant contribution to the torque integral arises on scales comparable to a secondary screening length of $O(Ri_v^{-\frac{1}{3}})$, and the resulting torque comes out to be $O(Ri_v^{\frac{2}{3}})$! Thus, the above arguments above show that, for small $Pe$, in addition to the $O(Ri_v)$ torque contribution given by (\ref{smallPe:prolateaspectratio}) and (\ref{smallPe:oblateaspectratio}), there exists a second independent contribution that increases with $Pe$ as $O(Ri_v^{\frac{1}{4}}Pe^{\frac{5}{4}})$ for $Pe  \ll Ri_v^{\frac{1}{3}}$, but is independent of $Pe$, being $(Ri_v^{\frac{2}{3}})$ for $Ri_v^{\frac{1}{3}} \ll  Pe  \ll 1$. This far-field hydrodynamic contribution, arising from a weak convective distortion of the stratified ambient, can evidently exceed the hydrostatic contribution, possibly leading to an edgewise settling regime for small $Pe$. In light of this additional contribution, the dominance of the $O(Ri_v)$ hydrostatic contribution and the prevalence of broadside-on settling requires the stricter criterion $Pe \ll  Ri_v^{\frac{3}{5}}$, instead of $Pe \ll 1$, as originally assumed. We will assume the former stricter criterion to hold. The far-field torque contribution above, along with a numerical investigation of the stratification torque over the entire range of $Pe$, will be reported in a separate communication.

\subsection{The angular velocity due to the hydrodynamic torque in the convection-dominant limit\,($Pe \gg 1$)} \label{sec:largePe}

In contrast to the small $Pe$ limit examined in the previous section, for $Pe \gg 1$, the dominant contribution to the integral for the stratification-induced hydrodynamic torque in (\ref{recithm_Angvelf}) comes from length scales of $O(LRi_v^{-\frac{1}{3}})$, the stratification screening length in the Stokes stratification regime\,($Re \ll Ri_v^{\frac{1}{3}}\ll 1$). To see this, we note from the right hand side of the advection diffusion equation (\ref{phys:densitycons}) that, for large $Pe$, the density perturbation is driven by the convection of the base-state stratification of order unity by the vertical component of the Stokeslet field, $u'^{(1)}_3$. Since $u'^{(1)}_3 \sim O(1/r)$, one finds $\rho'^{(1)} \sim O(1)$ for $r\gg 1$. This, along with the far-field $O(1/r^2)$ decay of the test velocity field, implies that the integrand in the stratification torque integral decays as $O(1/r^2)$, and that the volume integral therefore diverges as $O(r)$. This divergence is expected to be resolved only when the slow $O(1/r)$ decay of the Stokeslet is accelerated by stratification that, for large $Pe$, occurs on length scales of $O(LRi_v^{-\frac{1}{3}})$. It has recently been shown in \cite{Arun_2020} that, for a sedimenting sphere at large $Pe$, the density and velocity fields are indeed asymptotically small on length scales of $O(Ri_v^{-\frac{1}{3}})$, except within a horizontal wake whose vertical extent grows as $O(r_t^{\frac{2}{5}})$, $r_t$ being the distance in the plane transverse to gravity, where the density and axial velocity perturbation decay as $r_t^{-\frac{12}{5}}$ and $r_t^{-\frac{14}{5}}$, respectively; and a buoyant jet in the rear where $u_3$ reverses sign, but continues to exhibit an $O(1/r)$ Stokesian decay. Despite the latter slow decay, the asymptotically narrow character of the buoyant jet implies that the torque integral does converge on length scales of $O(LRi_v^{-\frac{1}{3}})$, and is $O(Ri_v^{-\frac{1}{3}})$. The pre-factor of $Ri_v$ in front of the integral in (\ref{recithm_Angvelf}) implies that the torque and the angular velocity scale as $O(Ri_v^{\frac{2}{3}})$ for $Pe \rightarrow \infty$. 

Since the dominant contribution to the torque integral comes from length scales much larger than $O(L)$, of $O(LRi_v^{-\frac{1}{3}})$, the calculation requires one to rewrite the integral involving $\rho'^{(1)}$, in (\ref{recithm_Angvelf}), in outer coordinates\,(defined below), with the sedimenting spheroid in problem 1 now regarded as a point force, and the rotating spheroid in the test problem acting as a combination of rotlet and stresslet singularities\,(see \cite{Marath_Timeperiod2017}). The details of this calculation are provided below, but before this it is worth noting that a reciprocal theorem formulation to determine the stratification-induced correction to the force\,(that would include both drag and lift components for an arbitrarily oriented spheroid) would involve the test problem of a translating spheroid instead. Since the test velocity field now decays as $O(1/r)$ in the far-field, this would lead to a stronger divergence of the force integral, as $O(Ri_v^{-\frac{2}{3}})$, and thence, an $O(Ri_v^{\frac{1}{3}})$ stratification-induced correction to the Stokes drag. Such a correction was originally calculated for a spherical particle by \cite{zvirin_settling_1975}. In a recent effort, \cite{dandekar_2020} have examined the force and torque acting on an arbitrarily shaped particle sedimenting in a linearly stratified ambient. For anisotropic particles lacking a handedness\,(that includes the spheroids examined here), the authors find a correction to the force at $O(Ri_v^{\frac{1}{3}})$ similar to the case of a spherical particle, but end up not finding a torque at this order, a result that is not surprising in light of the above scaling arguments which show the torque to be $O(Ri_v^{\frac{2}{3}})$. Within the framework of the matched asymptotics expansions approach used by the said authors, the $O(Ri_v^{\frac{1}{3}})$ correction to the drag appears as a response of the particle to an `ambient uniform flow' that is the limiting form of the outer solution in the matching region\,($1 \ll r \ll Ri_v^{-\frac{1}{3}}$); the uniformity of this flow is consistent with the absence of a torque at this order.

As mentioned above, the stratification torque integral in (\ref{recithm_Angvelfinal}) needs to be evaluated in outer coordinates which are related to the coordinates in the particle-fixed reference frame as $\tilde{\boldsymbol x} = Ri_v^{\frac{1}{3}}{\boldsymbol x}$, so an $O(1)$ change in $\tilde{\boldsymbol x}$ corresponds to ${\boldsymbol x}$ changing by an amount of order the stratification screening length. However, as originally shown by \cite{childress_1964} and \cite{saffman_1965}, a Fourier space approach turns out to be much more convenient for a calculation involving the outer region, and we therefore consider the Fourier transformed equations of continuity, motion and the advection diffusion equation for the density field, which are given by:
\begin{align}
k_i \hat{u}'^{(1)}_i =&\, 0, \label{ft:cont} \\
-4\pi^2 k^2\hat{u}'^{(1)}_i - 2\pi \mathrm{i}k_i \hat{p}'^{(1)} =&\,  - Ri_v(\hat{\rho}'^{(1)} \hat{g}_i + \tilde{F}\hat{g}_i), \label{ft:eqmot} \\
2\pi \mathrm{i}k_j \hat{U}_j \hat{\rho}'^{(1)} =&\, \hat{u}'^{(1)}_j\hat{g}_j + \frac{1}{Pe}4\pi^2 k^2 \hat{\rho}'^{(1)},   \label{ft:advdiff}
\end{align}
for problem 1. Here, we have used the definition $\hat{f}({\boldsymbol k}) = \textstyle\int d{\boldsymbol x} e^{-2\pi \mathrm{i} {\boldsymbol k}\cdot{\boldsymbol x}} f({\boldsymbol x})$ for the Fourier transform, and the sedimenting spheroid has been replaced by a point force, $\tilde{\boldsymbol F}\delta({\boldsymbol x})$, on the right hand side of the physical space equations of motion viz.  (\ref{phys:eqmot}), with $\tilde{\boldsymbol F} = \tilde{F}\hat{\boldsymbol g}$, $\tilde{F}$ being the non-dimensional buoyant force\,(in units of $\mu UL$) exerted by the spheroid; the corresponding dimensional expression has been given in section \ref{sec:recithm}\,(given that $U$ is itself defined in terms of $F$, this non-dimensionalization merely amounts also to mulplication by $X_A$). Note that the inertial term in the original physical space equations, (\ref{phys:eqcont})-(\ref{phys:densitycons}), has now been omitted since, as already argued earlier, the leading $O(Re)$ inertial torque is dominated by the inner region, with the outer region contribution being only $O(Re^2\ln Ri_v^{-\frac{1}{3}})$\,(see end of section \ref{sec:inertiastatic:torque}), and not considered here.

Setting $Pe =\infty$ in (\ref{ft:advdiff}), one obtains:
\begin{align}
\hat{\rho}'^{(1)} =&\,  \frac{\hat{u}'^{(1)}_j \hat{g}_j}{2\pi \mathrm{i} (k_l \hat{U}_l)}.  \label{ft:density}
\end{align}
Using (\ref{ft:density}) in (\ref{ft:eqmot}), and operating on both sides with $(\delta_{ij} - \hat{k}_i\hat{k}_j)$ to eliminate the pressure field, one obtains:
\begin{align}
4\pi^2 k^2\hat{u}'^{(1)}_i  =&\, Ri_v \hat{g}_j(\delta_{ij} - \hat{k}_i\hat{k}_j)\frac{\hat{u}'^{(1)}_l \hat{g}_l}{2\pi \mathrm{i} (k_p \hat{U}_p)}  + \tilde{F} \hat{g}_j (\delta_{ij} - \hat{k}_i \hat{k}_j).
\end{align}
Contracting with $\hat{\boldsymbol g}$ gives:
\begin{align}
\hat{u}'^{(1)}_i\hat{g}_i =&\,\frac{\tilde{F}\hat{g}_i\hat{g}_j(\delta_{ij} - \hat{k}_i\hat{k}_j)}{\biggl[ 4\pi^2 k^2 - Ri_v\frac{\hat{g}_i \hat{g}_j(\delta_{ij} - \hat{k}_i\hat{k}_j)}{2\pi \mathrm{i} (k_l \hat{U}_l)}\biggr]},
\end{align}
which, on using in (\ref{ft:density}), yields the following final expression for $\hat{\rho}'^{(1)}$:
\begin{align}
\hat{\rho}'^{(1)}({\boldsymbol k}) =&\, \frac{\tilde{F}[1-(\hat{k}_i \hat{g}_i)^2]}{\{8\pi^3\mathrm{i}k^3(\hat{k}_i \hat{U}_i) - Ri_v[1-(\hat{k}_i \hat{g}_i)^2]\}}, \label{ft:densitypert}
\end{align}
which will be used in the Fourier-space torque integral that is defined below.

The test velocity field $u^{(2)}_i =U^{(2)}_{ij}\Omega^{(2)}_j$ satisfies the Stokes equations, with the rotating spheroid, on length scales of $O(LRi_v^{-\frac{1}{3}})$, acting as a force-dipole singularity that includes both stresslet and rotlet contributions. Thus, the equations of motion may be written in the form\,(see \cite{Marath_Timeperiod2017}):
\begin{align}
\frac{\partial ^2 u^{(2)}_i}{\partial x_j^2} - \frac{\partial p^{(2)}}{\partial x_i} =&\, S^{(2)}_{ij} \frac{\partial }{\partial x_j} [\delta({\boldsymbol x})], \label{eq:pt_forcedipole}
\end{align}
where 
\begin{align}
S^{(2)}_{ij} = B_1[(\epsilon_{ilm}\Omega^{(2)}_lp_m )p_j + (\epsilon_{jlm}\Omega^{(2)}_lp_m )p_i] + B_3 \epsilon_{ijk}\Omega^{(2)}_k \label{fdipole:coeff}
\end{align}
with
\begin{align}
B_1 =&\, \frac{8\pi}{\xi_0^3(-3\xi_0 + 3 \coth^{-1}\xi_0(1+\xi_0^2))}, \\
B_3 =&\,\frac{8\pi(1-2\xi_0^2)}{\xi_0^3(-3\xi_0 + 3 \coth^{-1}\xi_0(1+\xi_0^2))}.
\end{align}
There is an additional contribution that is neglected  in (\ref{eq:pt_forcedipole}), on account of the test spheroid rotating about an axis transverse to ${\boldsymbol p}$, that is, since ${\boldsymbol \Omega}^{(2)} \cdot {\boldsymbol p}=0$. The term proportional to $B_3$ in (\ref{fdipole:coeff}) is the rotlet singularity\,(due to transverse rotation), while that involving $B_1$ is the stresslet singularity. Thus, for $\xi_0 \rightarrow \infty$, $B_3 = -4\pi$ and $B_1$ is $O(1/\xi_0^2)$, consistent with a rotating sphere acting as a pure rotlet singularity; note that $B_3 =Y_C/2$, the latter being the resistance function mediating the torque-angular-velocity relation for transverse rotation defined earlier.

Fourier transforming (\ref{eq:pt_forcedipole}), and contracting with the projection operator $({\boldsymbol I} -\hat{\boldsymbol k}\hat{\boldsymbol k})$, one obtains:
\begin{align}
\hat{u}^{(2)}_i =&\, - \frac{\mathrm{i}}{2\pi k}\{ B_1[(\epsilon_{mqr}\Omega^{(2)}_qp_r )p_n + (\epsilon_{nqr}\Omega^{(2)}_qp_r )p_m] + B_3 \epsilon_{mnq}\Omega^{(2)}_q\}\hat{k}_n(\delta_{im} - \hat{k}_i\hat{k}_m),
\end{align}
so the second-order tensor ${\boldsymbol U}^{(2)}$ is given by:
\begin{align}
U^{(2)}_{ij}({\boldsymbol k}) =&\,-\frac{\mathrm{i}}{2\pi k}\{ B_1[(\epsilon_{mjr} p_r )p_n + (\epsilon_{njr}p_r )p_m] + B_3 \epsilon_{mnj}\}\hat{k}_n(\delta_{im} - \hat{k}_i\hat{k}_m). \label{testvel:fdipole}
\end{align}

Now, using the convolution theorem, the integral for the angular velocity contribution due to the hydrodynamic component of the stratification-induced torque in (\ref{recithm_Angvelf}) may be written as:
\begin{align}
Ri_v \displaystyle\int \rho'^{(1)} \hat{g}_j U^{(2)}_{ji} dV =&\, Ri_v \displaystyle\int  d{\boldsymbol k} \hat{\rho}'^{(1)}({\boldsymbol k}) \hat{g}_j U^{(2)}_{ji}(-{\boldsymbol k}),  \label{convol_thm} 
\end{align}
where, in applying the convolution theorem, we have assumed the volume integral on the left hand side of (\ref{convol_thm}) to extend over the entire domain, and thereby, neglected the $O(L^3)$ volume of the spheroid. Since the dominant contribution\ arises from length scales of $O(LRi_v^{-\frac{1}{3}})$, this neglect only amounts to an error of $O(Ri_v)$ in the torque integral, and $O(Ri_v^2)$ in the resulting angular velocity. 

Using (\ref{ft:densitypert}) and (\ref{testvel:fdipole}) in the Fourier space torque integral in (\ref{convol_thm}), and after some simplification which includes defining a rescaled wavevector $2\pi{\boldsymbol k}$, one obtains the angular velocity induced by the hydrodynamic stratification torque as:
\begin{align}
\Omega_i^{(1)hd} =&\,Ri_v\frac{\mathrm{i}\tilde{F}}{8\pi^3Y_C} \displaystyle\int d{\boldsymbol k} \frac{[1-(\hat{k}_x\hat{g}_x)^2]}{\{\mathrm{i}k^3(\hat{k}_y \hat{U}_y) - Ri_v[1-(\hat{k}_z \hat{g}_z)^2]\}k}\left[ B_1\left\{ (\epsilon_{irj}p_r\hat{g}_j)(\hat{k}_mp_m)+ (\epsilon_{irj}p_r\hat{k}_j)(\hat{g}_mp_m) \right. \right. \nonumber \\
&\left. \left. - 2  (\hat{k}_m\hat{g}_m)(\hat{k}_jp_j)(\epsilon_{irl}p_r\hat{k}_l) \right\} + B_3 \epsilon_{ijr}\hat{g}_j \hat{k}_r \right], \label{strattorq:integral}
\end{align}
the terms proportional to $B_1$ and $B_3$ being the stresslet and rotlet-induced torque contributions, respectively. Redefining the new wavevector to be $Ri_v^{-\frac{1}{3}}{\boldsymbol k}$, so it remains of order unity on length scales of order the stratification screening length\,(and thereby, pertains to the outer region in Fourier space), and considering only the real part of the integral above, one obtains:
\begin{align}
\Omega_i^{(1)hd} =&\,Ri_v^{\frac{2}{3}} \frac{\tilde{F}}{8\pi^3Y_C} \left[B_1 \displaystyle\int d{\boldsymbol k}\frac{[1-(\hat{k}_x\hat{g}_x)^2]k^2(\hat{k}_v \hat{U}_v)}{\{k^6(\hat{k}_y \hat{U}_y)^2 +[1-(\hat{k}_z \hat{g}_z)^2]^2\}}\left[ \left\{ (\epsilon_{irj}p_r\hat{g}_j)(\hat{k}_mp_m)+ (\epsilon_{irj}p_r\hat{k}_j)(\hat{g}_mp_m)   \right. \right. \right.  \nonumber\\
&\left. \left. \left. - 2  (\hat{k}_m\hat{g}_m)(\hat{k}_jp_j)(\epsilon_{irl}p_r\hat{k}_l)\right\} \right] + B_3 \displaystyle\int d{\boldsymbol k}\frac{[1-(\hat{k}_x\hat{g}_x)^2]k^2(\hat{k}_v \hat{U}_v)}{\{k^6(\hat{k}_y \hat{U}_y)^2 +[1-(\hat{k}_z \hat{g}_z)^2]^2\}}\epsilon_{ijr}\hat{g}_j \hat{k}_r \right], \label{strattorque:integral1}
\end{align}
where the angular velocity due to the hydrodynamic stratification torque finally comes out to be $O(Ri_v^{\frac{2}{3}})$, as anticipated by the scaling arguments above, and the convergent Fourier integrals in (\ref{strattorque:integral1}) are evaluated below; note that the imaginary part of (\ref{strattorq:integral}), neglected in (\ref{strattorque:integral1}), may be shown to equal zero by symmetry. Before evaluating the integrals above using a specific coordinate system, we note that the force-velocity relationship for a sedimenting spheroid, for the scalings used here, is given by:
\begin{align}
\hat{U}_i = [p_i p_j + \frac{X_A}{Y_A}(\delta_{ij} - p_i p_j)] \hat{g}_j. \label{FU:relation}
\end{align}
Defining the aspect-ratio-dependent resistance ratio $An(\kappa) = X_A/Y_A$, (\ref{FU:relation}) may be written as:
\begin{align}
\hat{U}_i = [(1-An)p_i p_j + An \delta_{ij}] \hat{g}_j. \label{Ug_relation}
\end{align}
where $An$ decreases monotonically from unity for a sphere to a minimum of $1/2$ for an infinitely slender prolate spheroid\,($\kappa \rightarrow \infty$); on the oblate side, $An$ increases from unity to a maximum of $3/2$ for a flat disk\,($\kappa \rightarrow 0$). 

To evaluate the above Fourier integrals, we choose a spherical coordinate system with its polar axis along $\hat{\boldsymbol U}$. Interestingly, a numerical evaluation in the alternate and perhaps more natural choice of a $\hat{\boldsymbol g}$-aligned coordinate system turns out to much more involved, with the individual integrals making up the torque diverging as $Pe^{\frac{1}{2}}$ in the limit $Pe \rightarrow \infty$, the divergence arising due to the buoyant jet mentioned above, and that corresponds to the region $\hat{\boldsymbol k} \cdot \hat{\boldsymbol U} =0, k \ll 1$ in Fourier space\,(see \cite{Arun_2020}). We have verified that the divergences of the individual contributions cancel out, and the total torque integral is nevertheless convergent and independent of $Pe$ for $Pe \rightarrow \infty$, matching the result obtained below in the ${\boldsymbol U}$-aligned system. In the latter coordinate system, $\hat{\boldsymbol U} = U_m {\boldsymbol 1}_U$  where, using (\ref{Ug_relation}), one finds $U_m = \cos^2 \psi + An^2 \sin^2 \psi$, $\psi$ being the angle between $\hat{\boldsymbol g}$ and ${\boldsymbol p}$ defined in earlier sections. The unit wavevector is given by $\hat{\boldsymbol k} = \sin \theta \cos \phi {\boldsymbol 1}_{U^{\perp^1}} + \sin \theta \sin \phi {\boldsymbol 1}_{U^{\perp^2}} + \cos \theta {\boldsymbol  1}_U$, with ${\boldsymbol 1}_{U^{\perp^1}}$ chosen to lie in the plane of sedimentation, that is, the plane containing the vectors ${\boldsymbol p}$, $\hat{\boldsymbol U}$ and $\hat{\boldsymbol g}$, so the torque acting on the spheroid points along ${\boldsymbol 1}_{U^{\perp^2}}$ . With this choice, $\hat{\boldsymbol k} \cdot \hat{\boldsymbol U} =U_m \cos \theta$. We take ${\boldsymbol p} = \cos \psi_U {\boldsymbol 1}_U + \sin \psi_U {\boldsymbol 1}_{U^{\perp^1}}$, $\psi_U$ being the angle between ${\boldsymbol p}$ and $\hat{\boldsymbol U}$. Noting from (\ref{Ug_relation}) that $\hat{\boldsymbol U} \cdot {\boldsymbol p} = \hat{\boldsymbol g} \cdot {\boldsymbol p}$, one has $\cos \psi_U = \cos \psi/U_m$. With $\psi_U$ as defined above, $\hat{\boldsymbol k} \cdot {\boldsymbol p} = \cos \psi_U \cos \theta + \sin \psi_U \sin \theta \cos \phi$, and $\hat{\boldsymbol k} \cdot \hat{\boldsymbol g} = An^{-1}[\hat{\boldsymbol k} \cdot \hat{\boldsymbol U}  - (1-An) \cos \psi \hat{\boldsymbol k} \cdot {\boldsymbol p}]$. As mentioned above, the torque only has a component along ${\boldsymbol 1}_{U^{\perp^2}}$ in the chosen coordinate system. Since the vectors involved in the integrals in (\ref{strattorque:integral1}) are $\hat{\boldsymbol g} \wedge \hat{\boldsymbol k}$, ${\boldsymbol p} \wedge \hat{\boldsymbol g}$ and ${\boldsymbol p} \wedge \hat{\boldsymbol k}$, we have:
\begin{align}
\epsilon_{2jr}\hat{g}_j\hat{k}_r =&\,\frac{1}{An}[U_m \sin \theta \cos \phi - (1-An)\cos \psi(\cos \psi_U \sin \theta \cos \phi -  \sin \psi_U \cos \theta)], \label{int:exp1} \\
\epsilon_{2jr}p_j\hat{g}_r =&\,-\frac{U_m\sin\psi_U}{An}, \label{int:exp2} \\
\epsilon_{2jr}p_j\hat{k}_r =&\, - \sin \psi_U \cos \theta + \cos \psi_U \sin \theta \cos \phi, \label{int:exp3}
\end{align}
which defines all quantities involved in the calculation. Using these expressions in the integrals in (\ref{strattorque:integral1}), the $k$-integration is carried out analytically while the remaining two angular integrals are evaluated numerically using Gaussian quadrature. From (\ref{int:exp1}-\ref{int:exp3}), and other quantities defined in the preceding text, the $\psi$-dependence of the large-$Pe$ hydrodynamic angular velocity is seen to be more complicated than the $\cos \psi \sin \psi$ dependence obtained earlier for the inertial and hydrostatic contributions in section \ref{sec:inertiastatic:torque}, and for the hydrodynamic component, in the limit $Pe \ll 1$, in section \ref{sec:smallPe}. This is along expected lines since the large $Pe$ limit examined in this section is a singular perturbation problem, as evident from the outer region contributing at leading order. Another feature of the singular character is that, unlike the earlier torque contributions, the large-$Pe$ hydrodynamic stratification torque is in general a non-separable function of $\psi$ and $\kappa$.

\begin{figure}
\centering
\begin{subfigure}{\textwidth}
  \centering
  \includegraphics[width=.8\linewidth]{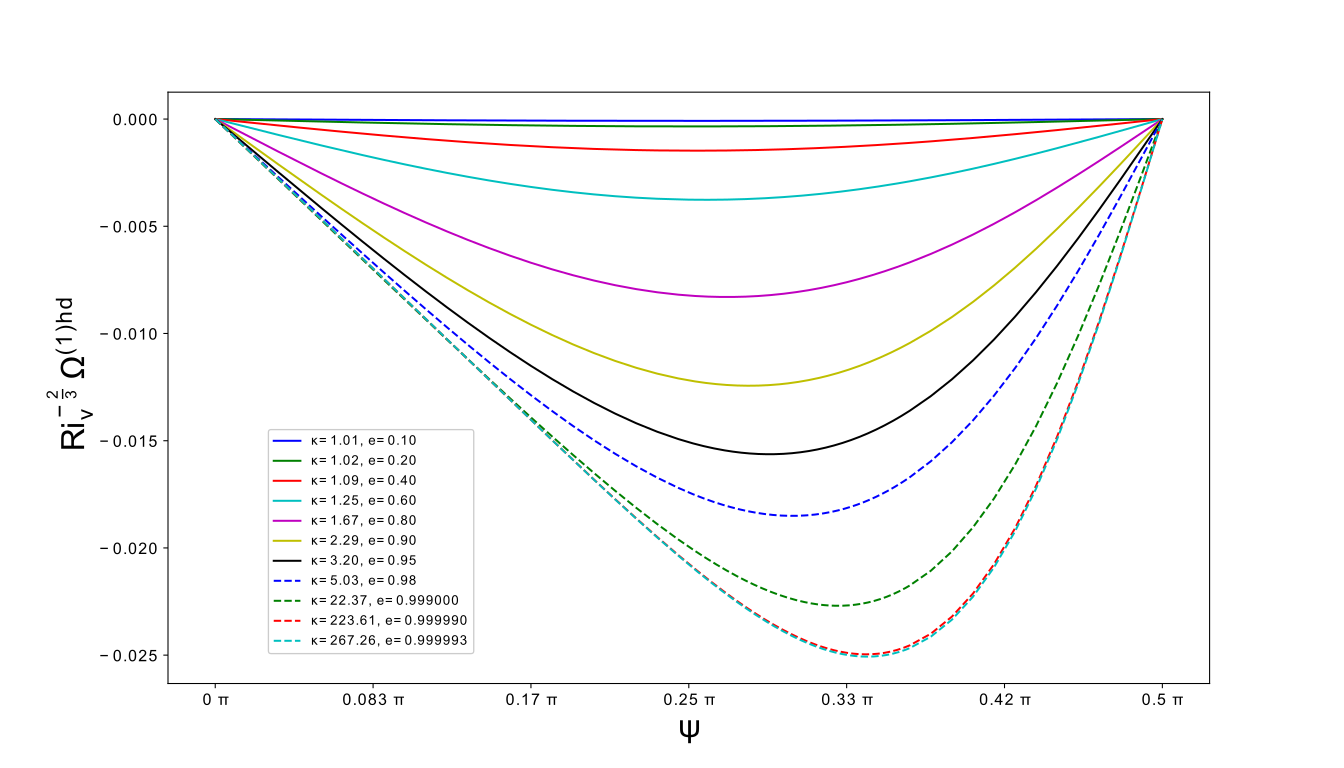}
  \caption{ $\Omega^{(1)hd}$ for prolate spheroids}
  \label{fig:InfPeProlate}
\end{subfigure}\\%\\
\begin{subfigure}{\textwidth}
  \centering
  \includegraphics[width=.8\linewidth]{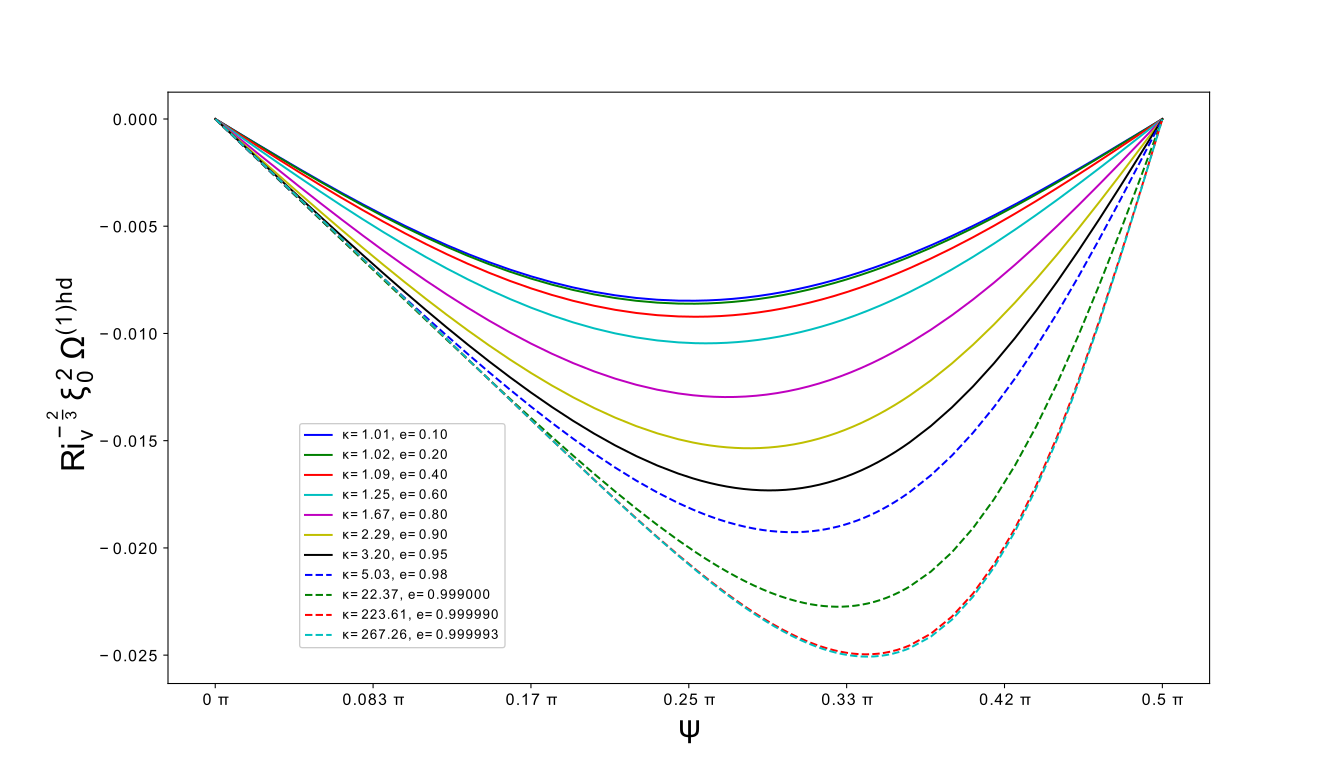}
  \caption{ $\Omega^{(1)hd}$ scaled with its near sphere scaling}\label{fig:InfPeProlateNS}
  
\end{subfigure}\\

\caption{Angular velocity due to the hydrodynamic component of the stratification torque for prolate spheroids}
\label{fig:test}
\end{figure}

\begin{figure}
\centering
\begin{subfigure}{\textwidth}
  \centering
\includegraphics[width=.8\linewidth]{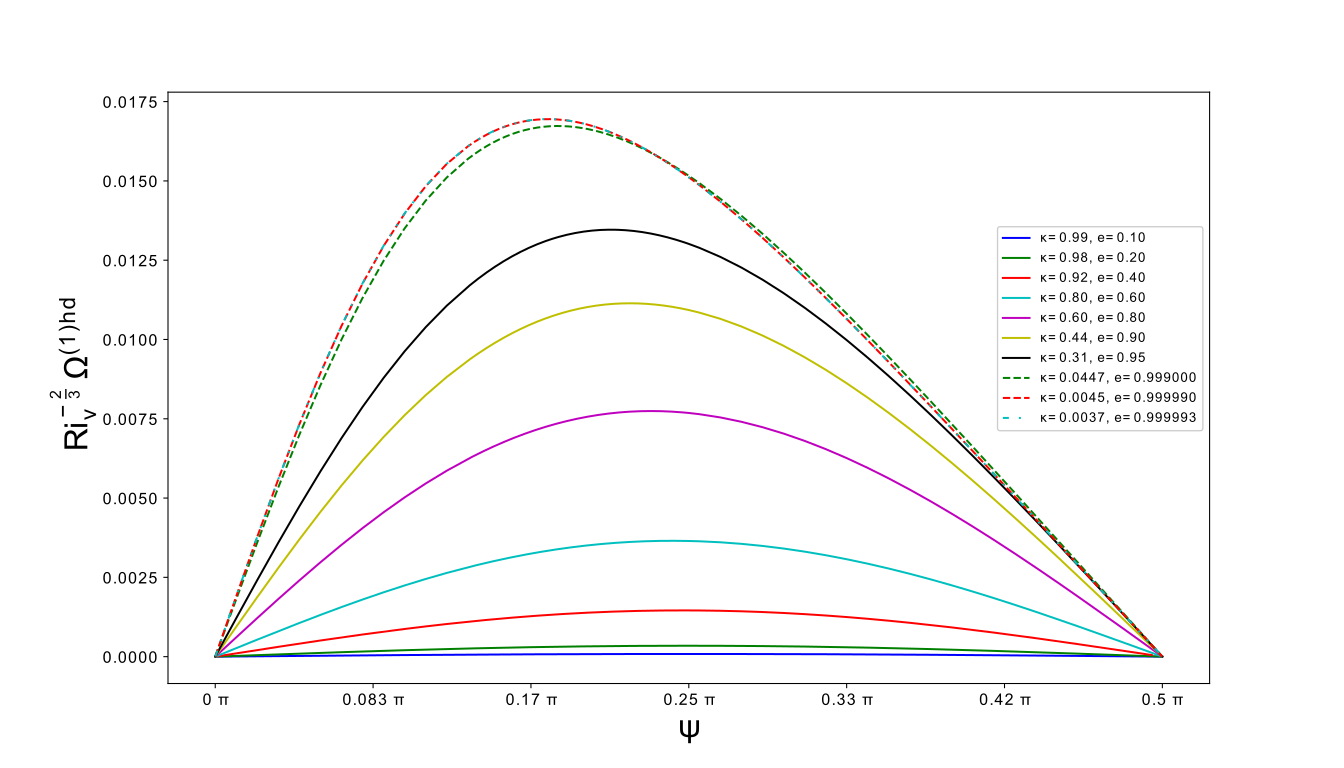}
\caption{$\Omega^{(1)hd}$ for oblate spheroids}\label{fig:InfPeOblate}
\end{subfigure}\\%\\
 
\begin{subfigure}{\textwidth}
  \centering
\includegraphics[width=.8\linewidth]{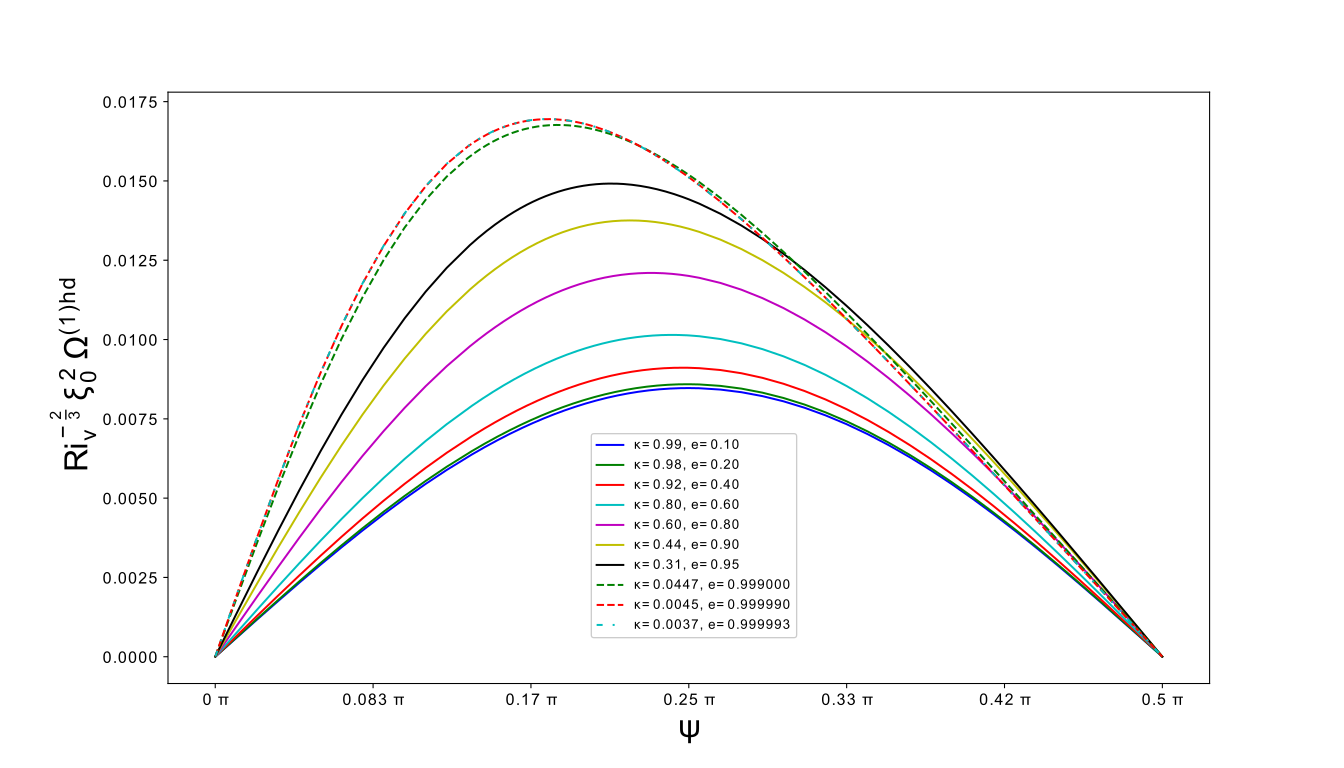}
\caption{$\Omega^{(1)hd}$ scaled with its near sphere scaling}\label{fig:InfPeOblateNS}
  
\end{subfigure}\\
\caption{Angular velocity due to the hydrodynamic component of the stratification torque for oblate spheroids}
\end{figure}
\begin{figure}
\centering
\includegraphics[width=\linewidth]{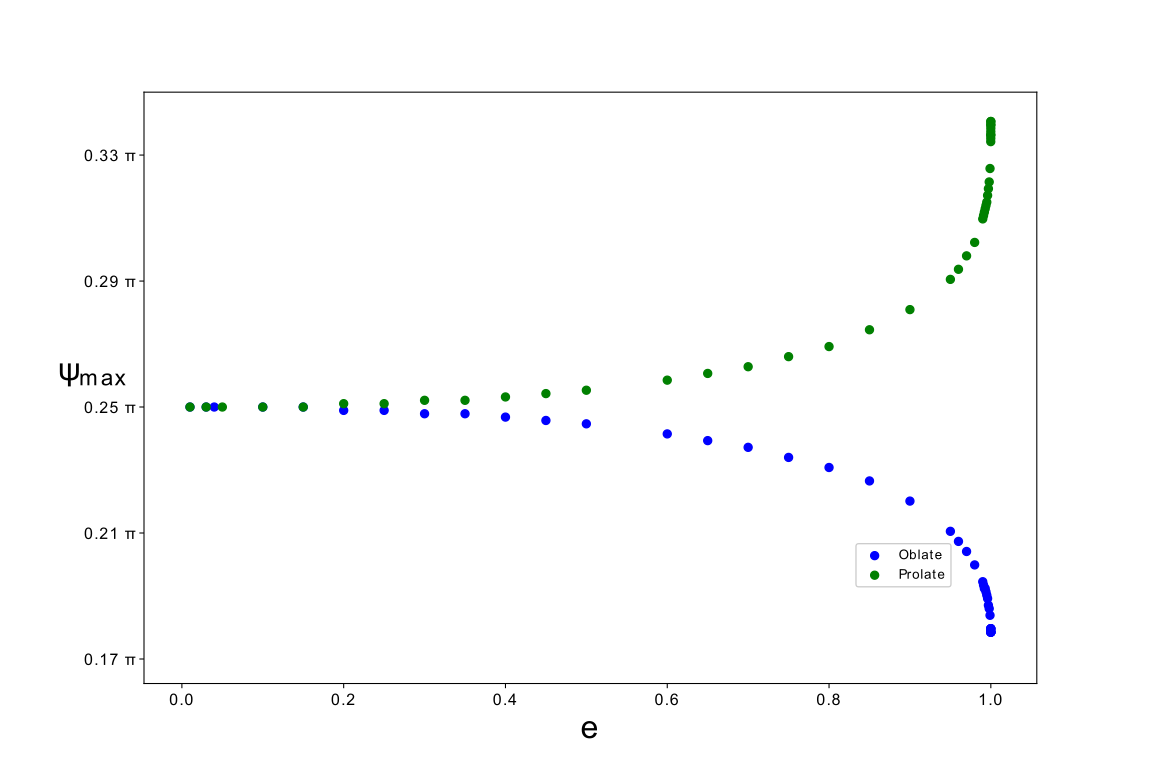}
\caption{The angle corresponding to the maximum angular velocity, arisng from the hydrodynamic component of the stratification torque, plotted as a function of the spheroid aspect ratio\,(both prolate and oblate spheroids).}\label{fig:MaxPsi}
\end{figure}

\begin{figure}
\centering
\begin{subfigure}{\textwidth}
  \centering
\includegraphics[width=1.2\linewidth]{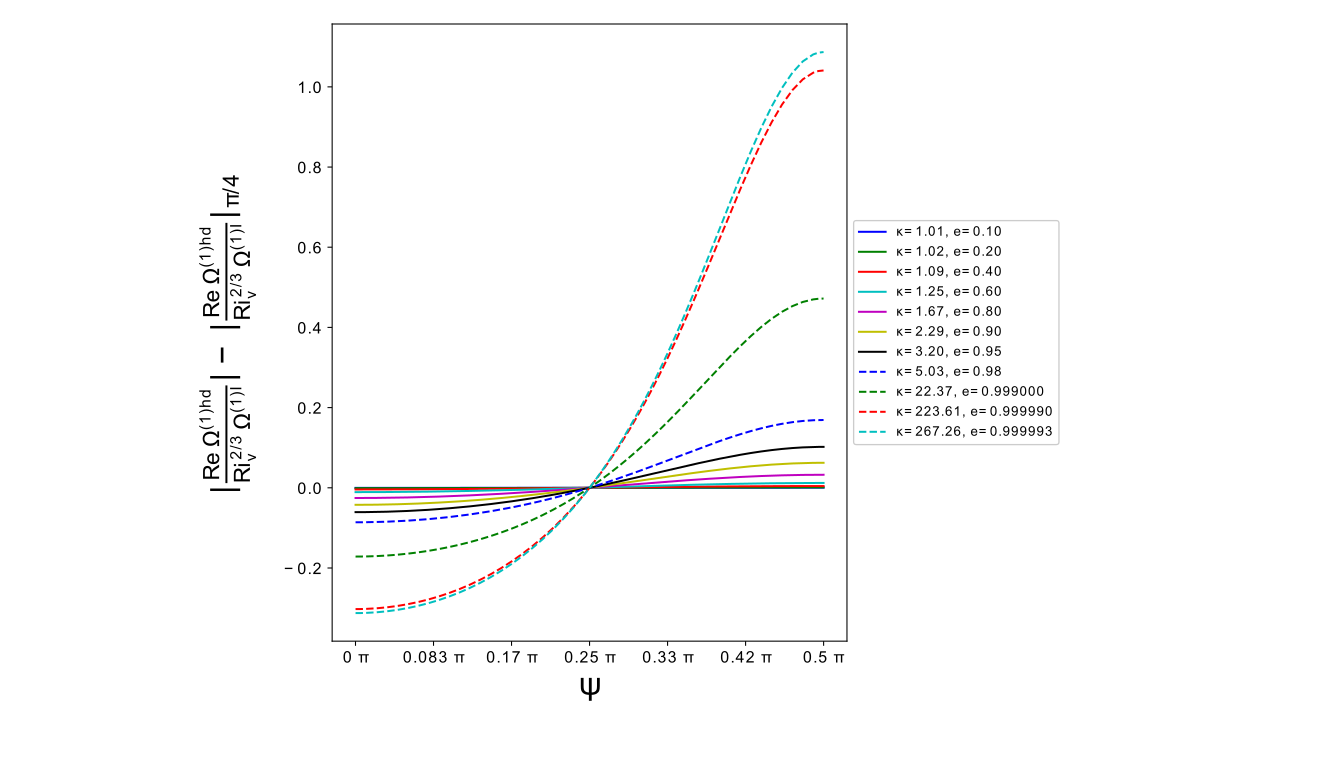}
\caption{Prolate Spheroid}\label{fig:InfPeProlateRatio}
\end{subfigure}\\%\\
\begin{subfigure}{\textwidth}
  \centering
\includegraphics[width=1.2\linewidth]{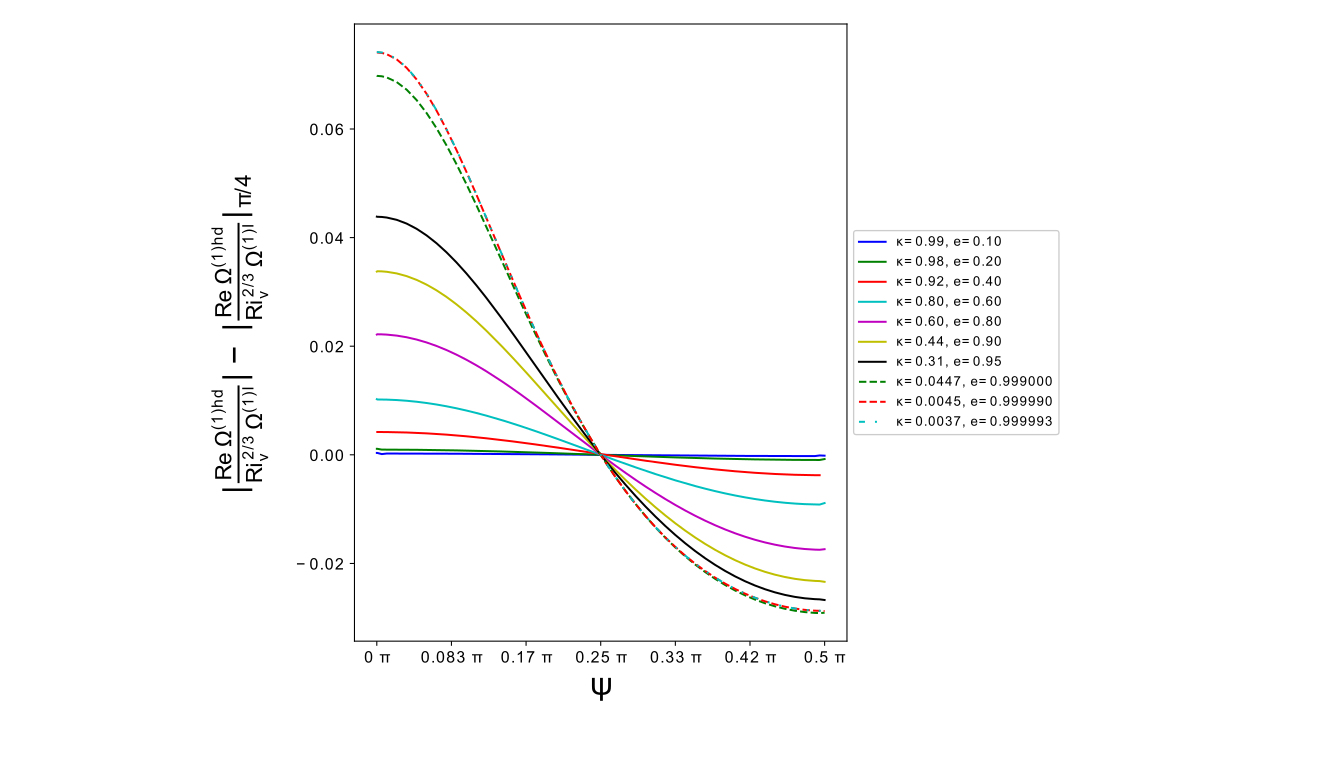}
\caption{Oblate Spheroid}\label{fig:InfPeOblateRatio}
\end{subfigure}\\
\caption{Ratio of the angular velocities due to the hydrodynamic-stratification and inertial torques. The stratification-induced rotation is higher for near-vertical and near-horizontal orientations for prolate and oblate spheroids, respectively}\label{fig:Ratio}
\end{figure}

Figures \ref{fig:InfPeProlate} and \ref{fig:InfPeOblate} show plots of the angular velocity\,($\Omega^{(1)hd}$), due to the hydrodynamic component of the stratification torque, versus $\psi$  for prolate and oblate spheroids, respectively. As evident from Figure \ref{fig:InfPeProlate}, for prolate spheroids, the magnitude of the angular velocity is expectedly small in the near-sphere limit, increasing monotonically with $\kappa$ to a maximum in the limit of a slender fiber. Figures \ref{fig:InfPeProlateNS} shows plots of the angular velocity scaled with the square of the eccentricity\,($\xi_0^{-2}$), so as to obtain a collapse in the near-sphere limit. Note that the finite value of the stratification-induced angular velocity in the limit of a slender fiber\,($\xi_0 \rightarrow 1$) is in contrast to the $O[\ln (\xi_0-1)]^{-1}$ scaling exhibited by the inertial angular velocity calculated in section \ref{sec:inertiastatic:torque}\,(also see \cite{Marath_JFM2015}), and implies that, for fixed $Re$ an $Ri_v$, the stratification torque invariably becomes dominant for large aspect ratios. Figure \ref{fig:InfPeOblateNS} confirms the squared-eccentricity scaling for oblate spheroids with near-unity aspect ratios; expectedly, the angular velocity approaches a finite value in the limit of a flat disk. For both prolate and oblate spheroids, the sign of $\Omega^{(1)hd}$ is such as to rotate the spheroid onto an edgewise orientation. The non-trivial orientation dependence of the angular velocity referred to in the previous paragraph is also evident from the plots in figures \ref{fig:InfPeProlate} and \ref{fig:InfPeOblate}. For near-unity aspect ratios, the angular velocity curve is nearly symmetric about $\psi = \frac{\pi}{4}$; that the angular dependence in this limit is indeed of the form $\sin \psi \cos \psi$ may be shown based on the fact that for $An \rightarrow 1$, $\psi_U \approx \psi$. The asymmetry about $\psi = \frac{\pi}{4}$ increases as the aspect ratio departs from unity, with the location of the maximum angular velocity moving to $\psi$'s greater than, and less than, $\frac{\pi}{4}$ for prolate and oblate spheroids, respectively, as shown in figure \ref{fig:MaxPsi}. To see the deviation of the angular dependence from the aforementioned simple form more clearly, in figures \ref{fig:InfPeProlateRatio} and \ref{fig:InfPeOblateRatio} we plot the angular velocity, scaled by the inertial angular velocity which is proportional to $\sin \psi \cos \psi$, again as a function of $\psi$. For near-unity aspect ratios, one obtains a horizontal line, while for both larger and smaller aspect ratios, this renomalized angular velocity asymptotes from one plateau for $\psi \rightarrow 0$ to a second for $\psi \rightarrow \frac{\pi}{2}$.

In contrast to the inertial contribution determined in section \ref{sec:inertiastatic:torque}, which was independent of the ambient stratification at leading order, the stratification-induced torque can, in principle, be coupled to inertial forces even in the limit $Re, Ri_v \ll 1$. For sufficiently small $Pe$\,($Pe \ll Ri_v^{\frac{3}{5}}$ as argued in section \ref{sec:smallPe}), the density perturbation that determines the stratification torque arises from a diffusive response to the no-flux condition on the surface of the sedimenting spheroid, and is therefore independent of the fluid motion. As a result, a non-trivial coupling between stratification and inertia can occur only for large $Pe$. Since the dominant length scales contributing to the stratification torque in this limit are much larger than $O(L)$, the magnitude of the density perturbation is controlled by the convection of the ambient stratification by the far-field disturbance fluid motion. The nature of this convection is therefore dependent on the form of the disturbance velocity field, and this in turn depends on the relative magnitudes of the inertial\,($LRe^{-1}$) and stratification\,($LRi_v^{-\frac{1}{3}}$) screening lengths. The calculation of the angular velocity due to the hydrodynamic stratification torque above pertains to the Stokes stratification regime, with $Re \ll Ri_v^{\frac{1}{3}}$, where the disturbance velocity field directly transitions from the Stokeslet form to a more rapid decay on length scales of $O(LRi_v^{-\frac{1}{3}})$\,(\cite{Arun_2020}) and is therefore independent of $Re$. In the Stokes stratification regime therefore, the stratification-induced rotation, both in the limit of small and large $Pe$, is independent of $Re$, and the inertial and stratification angular velocity contributions are additive. This will no longer true when $Re \geq O(Ri_v^{\frac{1}{3}})$, corresponding to the so-called inertia-stratification regime, in which case the leading order stratification-induced rotation for large $Pe$ will be a function of $Re/Ri_v^{\frac{1}{3}}$. In the limit $Ri_v^{\frac{1}{3}} \ll Re \ll 1$, opposite to the one analyzed above, the disturbance velocity transitions from an $O(1/r)$ to an $O(1/r^2)$ decay\,(outside of a viscous wake) across length scales of order the inertial screening length. This leads to the stratification torque integrand decaying as $O(1/r^3)$ for length scales much larger than $O(LRe^{-1})$, and the torque integral in (\ref{recithm_Angvelfinal}) continues to exhibit a logarithmic divergence. This (milder)\,divergence is only eliminated when buoyancy forces become comparable to inertial forces at a secondary screening length that was estimated in \cite{Arun_2020} to be $O(Re/Ri_v)^{\frac{1}{2}}$. Accounting for the aforementioned cut-off of the logarithmic divergence, the angular velocity arising from the hydrodynamic stratification torque is expected to have a leading $O[Ri_v Re^{-1}\ln(Re/Ri_v^{\frac{1}{3}})]$ contribution arising from a region between the primary and secondary screening lengths\,(that is, due to the logarithmic growth for $Re^{-1} \ll r \ll (Re/Ri_v)^{\frac{1}{2}}$), with logarithmically smaller $O(Ri_v Re^{-1})$ contributions arising from length scales of order the two screening lengths. Assuming this angular velocity to rotate the spheroid towards an edgewise configuration, and equating it to the $O(Re)$ inertial contribution, one obtains $Re \leq Ri_v^2$ for a transition to an edgewise-settling regime. This, however, contradicts the requirement $Re \gg Ri_v^{\frac{1}{3}}$ characterizing the inertia-stratification regime, implying that the inertial angular velocity contribution remains dominant in this regime. Thus, in the limit of small $Re$ and $Ri_v$, a broadside-on-edgewise transition is possibly only in the Stokes stratification regime.

To end this subsection, we again examine the validity of a quasi-steady state assumed in the analysis above for large $Pe$. As already argued in section \ref{sec:inertiastatic:torque}, momentum diffusion occurs asymptotically fast for small $Re$, and therefore, the quasi-steady assumption used to evaluate the stratification torque integral relies on the time scale for the density disturbance to approach a steady state being much shorter than that characterizing spheroid rotation. The former time scale may be regarded as that required to convect the density perturbation through the $O(LRi_v^{-\frac{1}{3}})$ stratification screening length, and is therefore $O(L/URi_v^{-\frac{1}{3}})$. The time scale of rotation is $O(L/URe^{-1})$ or $O(L/URi_v^{-\frac{2}{3}})$, depending on which of $Re$ or $Ri_v^{\frac{2}{3}}$ is greater. In either case the time scale for the development of a steady density perturbation is smaller, provided one remains in the Stokes stratification regime $Re \ll Ri_v^{\frac{1}{3}}$. Note that the perturbation density field, for $Pe = \infty$, is expected to be logarithmically singular along the rear stagnation streamline, as has been shown for the case of a spherical particle\,(see \cite{Arun_2020}), with this singularity either being resolved on a larger diffusive time scale\,(that is, due to $Pe$ being regarded as large but finite), or on account of the unsteadiness arising from the rotation of the settling spheroid. However, the convergence of the Fourier integrals involved in the stratification torque above implies that the contribution of the transiently developing region in the immediate neighborhood of the rear stagnation streamlilne is irrelevant as far as the leading order hydrodynamic stratification torque is concerned, and a quasi-steady analysis of this torque remains valid for $Re \ll Ri_v^{\frac{1}{3}}$.

\section{Results and Discussion} \label{sec:results}
\begin{figure}
\centering
\begin{subfigure}{.5\textwidth}
  \centering
  \includegraphics[width=\linewidth]{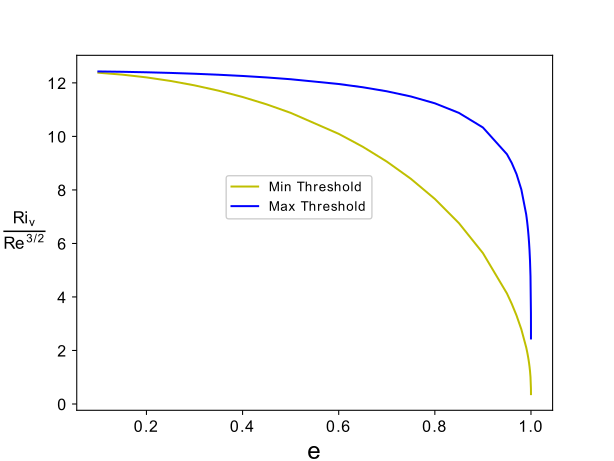}
  %\caption{A subfigure}
  \label{fig:sub1}
\end{subfigure}%
\begin{subfigure}{.5\textwidth}
  \centering
  \includegraphics[width=\linewidth]{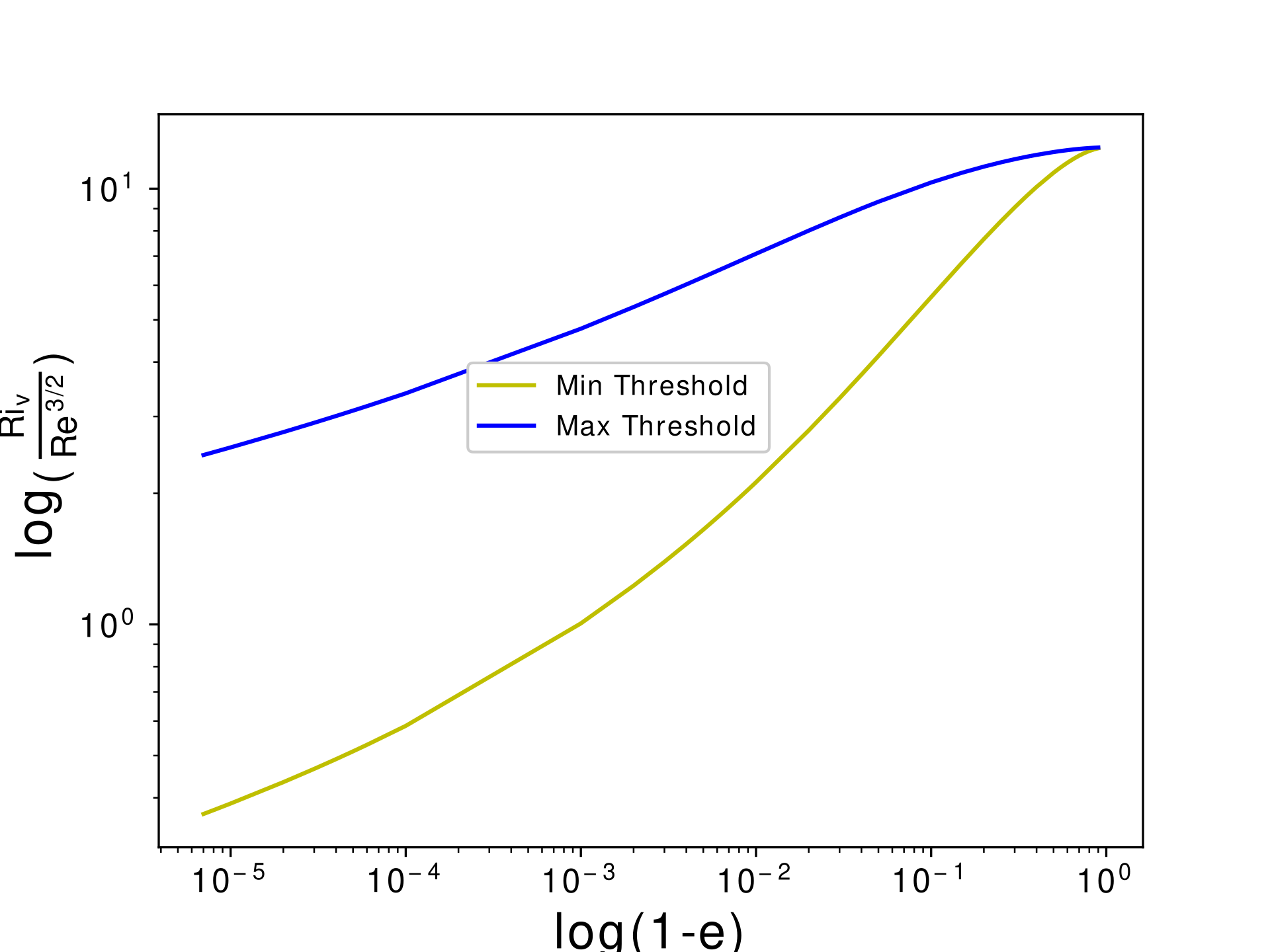}
  %\caption{A subfigure}
  \label{fig:sub2}
\end{subfigure}
\caption{The upper and lower threshold curves that demarcate the regimes of broadside-on settling\,(below), edgewise settling\,(above) and intermediate equilibrium orientations\,(in between), plotted as a function of eccentricity, for a prolate spheroid; the plot on the right presents a magnified view of the thresholds near the slender-fiber limit.}
\label{fig:thresholdprolate}
\end{figure}

\begin{figure}
\centering
\begin{subfigure}{.5\textwidth}
  \centering
  \includegraphics[width=\linewidth]{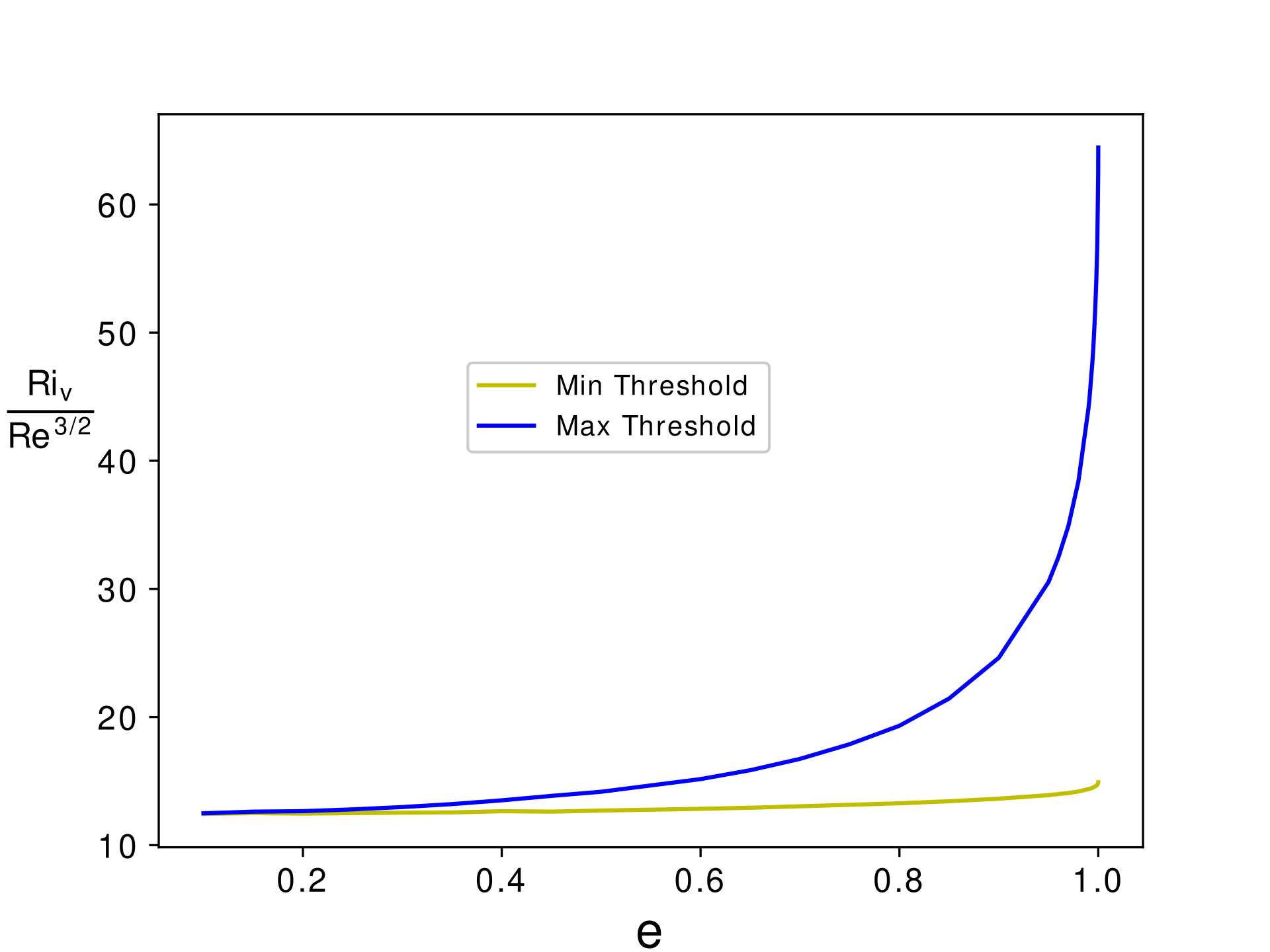}
  %\caption{A subfigure}
  \label{fig:sub1}
\end{subfigure}%
\begin{subfigure}{.5\textwidth}
  \centering
  \includegraphics[width=\linewidth]{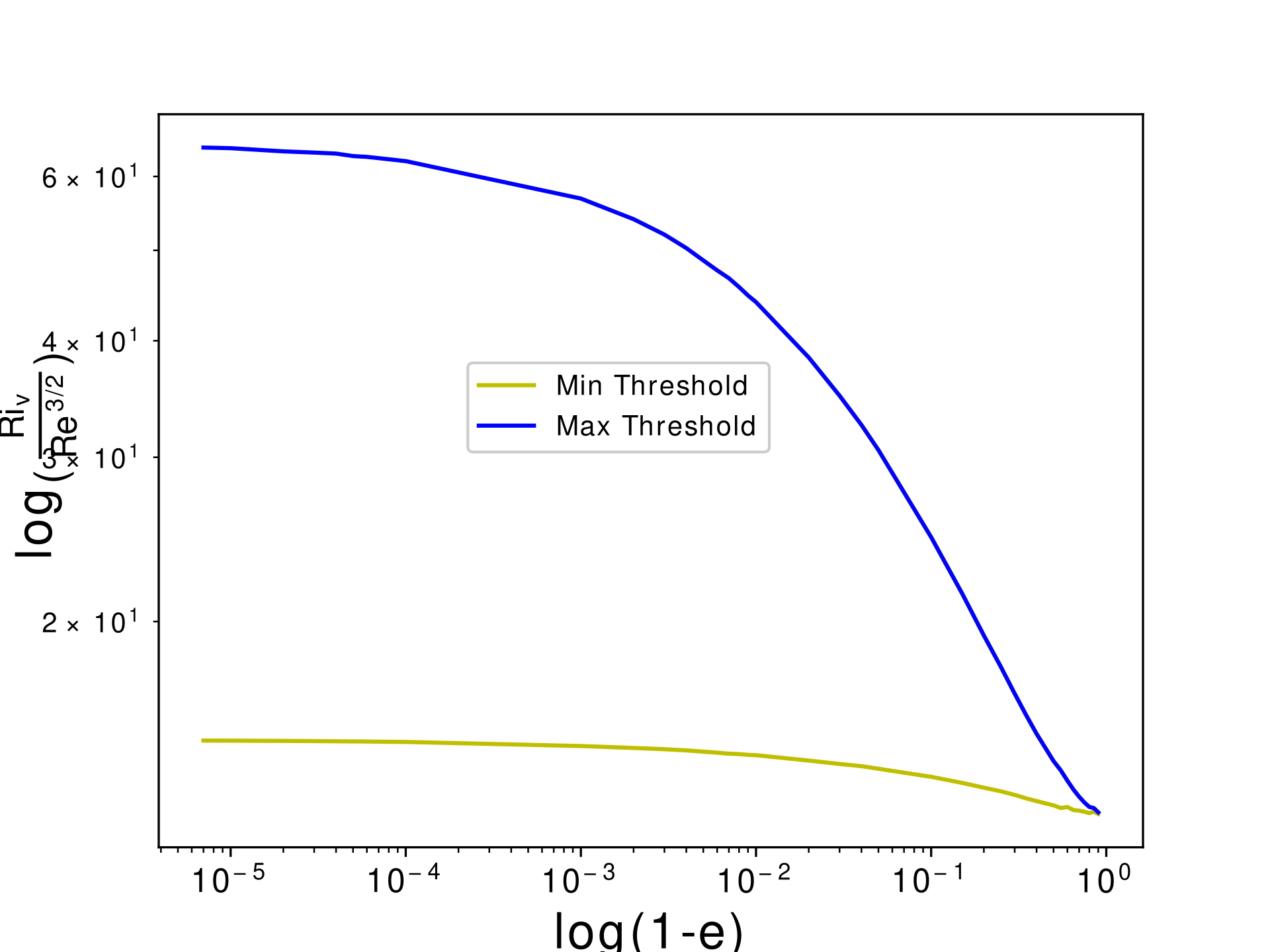}
  %\caption{A subfigure}
  \label{fig:sub2}
\end{subfigure}
\caption{The upper and lower threshold curves that demarcate the regimes of broadside-on settling\,(below), edgewise settling\,(above) and intermediate equilibrium orientations\,(in between), plotted as a function of eccentricity, for an oblate spheroid; the plot on the right presents a magnified view of the thresholds near the flat-disk limit.}
\label{fig:thresholdoblate}
\end{figure}
In earlier sections, we have derived expressions for the angular velocity of a spheroid settling in a viscous linearly stratified ambient. The spheroid angular velocity is the sum of three components; the inertial and hydrostatic contributions for a prolate spheroid are given by (\ref{eqn:prolate_inertialtorque}) and (\ref{eqn:prolate_hydrostattorque}) ((\ref{eqn:oblate_inertialtorque}) and (\ref{eqn:oblate_hydrostattorque})) for prolate (oblate) spheroids; the hydrodynamic contribution arising from the stratification is given by (\ref{smallPe:prolateaspectratio}) and (\ref{smallPe:oblateaspectratio}) for prolate and oblate spheroids, respectively, in the limit $Pe \ll 1$; and is obtained from the numerical evaluation of (\ref{strattorque:integral1}) for $Pe \gg 1$. As already argued in section \ref{sec:smallPe}, both prolate and oblate spheroids will settle broadside-on for sufficiently small $Pe$ regardless of $\kappa$. Herein, we therefore focus on the transition from broadside-on to edgewise settling that becomes possible for large $Pe$. In this limit, the hydrodynamic stratification component is $O(Ri_v^{\frac{2}{3}})$ and rotates the spheroid towards an edgewise orientation regardless of $\kappa$. It is dominant over the $O(Ri_v)$ hydrostatic component that favors the broadside-on orientation. Thus, the transition from broadside-on to edgewise settling depends on the relative magnitudes of the inertial and hydrodynamic stratification angular velocities, and for a fixed $\kappa$, the transition threshold is determined by the ratio $Ri_v/Re^{\frac{3}{2}}$ in the limit $Re, Ri_v \ll 1$. However, the differing orientation dependence of the inertial and stratification angular velocities, as evident from comparing figure \ref{fig:InfPeProlateRatio} for prolate spheroids, for instance, implies that the transition cannot be characterized by $Ri_v/Re^{\frac{3}{2}}$ equalling a single $\kappa$-dependent threshold. An instance of the latter scenario is when the competing physical effects are inertia and viscoelasticity, both of which exhibit a $\sin \psi \cos \psi$ dependence, so that the edgewise and broadside-on settling regimes are demarcated by a single critical curve in the $De/Re-\kappa$ plane, $De$ here being the Deborah number, a dimensionless measure of elasticity\,(see \cite{Marath_JFM2015}).

Writing the leading order hydrodynamic stratification component in the general form $Ri_v^{\frac{2}{3}}F_s(\kappa,\psi)$, for large $Pe$, and equating it to the inertial component, of the form $F_I(\kappa)\sin \psi \cos \psi$, the threshold criterion for the broadside-on-edgewise transition is determined by the ratio $Ri_v/Re^{\frac{3}{2}} = [(\sin \psi \cos \psi)F_I(\kappa)/F_s(\kappa,\psi)]^{\frac{3}{2}}$. Recall from figure \ref{fig:Ratio} that, for all $\kappa>1$\,($\kappa<1$), $F_s(\kappa,\psi)/(F_I(\kappa)\sin \psi \cos \psi)$ approaches its minimum and maximum values for $\psi \rightarrow 0\,(\pi/2)$ and $\pi/2\,(0)$, respectively, varying monotonically in between these limits. Now, define $(Ri_v/Re^{\frac{3}{2}})_{max} = \lim_{\psi \rightarrow 0(\psi \rightarrow\frac{\pi}{2})}[(F_I(\kappa)\sin \psi \cos \psi)/F_s(\kappa,\psi)]^{\frac{3}{2}}$ and $(Ri_v/Re^{\frac{3}{2}})_{min} = \lim_{\psi \rightarrow \frac{\pi}{2}(\psi \rightarrow 0)}[(F_I(\kappa)\sin \psi \cos \psi)/F_s(\kappa,\psi)]^{\frac{3}{2}}$ for prolate\,(oblate) spheroids, both of which are finite and only functions of $\kappa$. One then has the following behavior for the orientation of either a sedimenting prolate or an oblate spheroid. For $Ri_v/Re^{\frac{3}{2}} < (Ri_v/Re^{\frac{3}{2}})_{min}$, the broadside-on orientation is the only equilibrium; likewise, for $Ri_v/Re^{\frac{3}{2}} > (Ri_v/Re^{\frac{3}{2}})_{max}$ the longside-on orientation is the only equilibrium. For $Ri_v/Re^{\frac{3}{2}}$ between the aforementioned thresholds, the inertial and stratification angular velocity curves must intersect at an orientation, $\psi_i$\,(say), intermediate between $0$ and $\pi/2$. It is easily seen that this equilibrium is a stable one for both the prolate and oblate cases; for example, in the prolate case, the stratification-induced rotation is greater than the inertial one for $\psi_i < \psi <\pi/2$, with the converse being true $0 <\psi_i < \psi$, implying that a prolate spheroid with its orientation in either of these intervals is rotated towards $\psi = \psi_i$. As $Ri_v/Re^{\frac{3}{2}}$ increases from the lower to the upper threshold, the intermediate equilibrium orientation, $\psi_i$, decreases from $\frac{\pi}{2}$ to zero. Figures \ref{fig:thresholdprolate} and \ref{fig:thresholdoblate} show the aforementioned pair of threshold curves, $(Ri_v/Re^{\frac{3}{2}})_{min}(\kappa)$ and $(Ri_v/Re^{\frac{3}{2}})_{max}(\kappa)$, plotted in the $Ri_v/Re^{\frac{3}{2}}-\kappa$ plane for prolate and oblate spheroids, respectively. Both the threshold values in figure \ref{fig:thresholdprolate} approach zero in the limit of large aspect ratios because, as already noted in section \ref{sec:largePe}, the stratification-induced torque remains finite in this limit, in contrast to the inertial torque which becomes logarithmically small\,(see figure \ref{fig:inertia}). As seen from the log-log plot in figure \ref{fig:thresholdprolate}, the convergence of the thresholds to zero is slow on account of the aforementioned logarithmic scaling. For oblate spheroids, the lower and upper thresholds approach distinct finite values in the limit of a flat disk. Since the angular velocity due to the hydrodynamic stratification torque approaches a $\sin \psi  \cos \psi$ dependence for $\kappa \rightarrow 1$ from either the prolate or oblate side, the two threshold curves towards a common albeit finite critical value in the near-sphere limit in both figures \ref{fig:thresholdprolate} and \ref{fig:thresholdoblate}. In effect, for a prolate spheroid, the thresholds diverge from a common finite value for $\kappa = 1$, tending to a maximum separation for $\kappa \approx 4.11$, before approaching zero in the limit $\kappa \rightarrow \infty$. For flat disks, the threshold curves diverge away monotonically from a common value as $\kappa$ increases from unity, approaching a maximum separation in the limit of a flat disk.

In order to connect to experiments, we now discuss the implication of the aforementioned predictions, within a quasi-steady framework, for a spheroid that starts off with an arbitrary initial orientation and sediments through a stratified fluid at large $Pe$; arguments in sections \ref{sec:smallPe} and \ref{sec:largePe} show that the quasi-steady assumption remains rigorously valid in the Stokes stratification regime, regardless of $Pe$. The experiments reported in \cite{mercier_2020} correspond to an ambient linear stratification that includes a neutral buoyancy level. The latter would correspond to the equilibrium location of the sedimenting spheroid for long times, and for the viscous overdamped regime under consideration, one expects the spheroid velocity $U$ to decrease monotonically to zero as it approaches this level. In dimensionless terms, $Re$ and $Pe$ decrease with time, while $Ri_v$ increases with time. If the spheroid starts off sufficiently far above neutral buoyancy level, then the initial terminal velocity is likely large enough for the ratio $Ri_v/Re^{\frac{3}{2}} \sim U^{-\frac{5}{2}}$ to be below the lower $\kappa$-dependent threshold in figure \ref{fig:thresholdoblate}, or strictly speaking, the finite-$Re-Ri_v$ analog of this threshold\,(note that the particles used in the experiments were disk-shaped, and maybe likened to thin oblate spheroids). As a result, the spheroid starts off rotating towards a broadside-on orientation. The spheroid will slow down as it approaches the neutral buoyancy level, and the resulting increase in $Ri_v/Re^{\frac{3}{2}}$ will eventually cause it to exceed the aforementioned lower threshold, leading to the broadside-on orientation becoming an unstable equilibrium. Assuming the spheroid to have had sufficient time prior to this point, to have already attained a near-broadside-on orientation, one expects the onset of a reversal in rotation. Strictly speaking, the arguments in the previous paragraph, with regard to the existence of an intermediate stable equilibrium, only pertain to a truly steady setting\,(where the neutral buoyancy level corresponds to an infinitely great depth). For the experimental scenario, assuming a sufficiently slow decrease in $Ri_v/Re^{\frac{3}{2}}$ with time, the spheroid would progress quasi-statically through a sequence of intermediate orientation equilibria, on its way to an edgewise configuration. Finally, in the immediate neighborhood of the neutral buoyancy level, the dynamics is slow enough for one to be in the small-$Pe$ regime analyzed in section \ref{sec:smallPe}, and the resulting dominance of the hydrostatic component of the stratification torque, over the $O(Ri_v)$ hydrodynamic component, should again reverse the spheroid rotation, causing it to finally approach its equilibrium location in a broadside-on configuration. The aforementioned sequence of events is broadly consistent with the observations in \cite{mercier_2020}. Note that $U\rightarrow 0$ for long times, and this leads to $Ri_v$ becoming arbitrarily large in the vicinity of the neutral buoyancy level, in turn leading to an apparent breakdown of the analysis in section \ref{sec:smallPe} that predicts the second and final reversal in rotation towards a broadside-on orientation above. However, as pointed out therein, this breakdown is only an apparent one and arises due to the use of $a/U$ as the angular velocity scale which is inappropriate for small $Pe$ when the dimensional angular velocity is independent of $U$, and $O(\gamma L^2 g/\mu)$.

The experiments reported in \cite{Mrokowska_2018},\cite{Mrokowska_20201} and \cite{Mrokowska_20202}) correspond to a non-linearly stratified ambient where the density varies within an intermediate layer sandwiched between homogeneous upper and lower layers. The effects of the stratification on particle orientation, and the resulting coupling to the settling velocity via the orientation-dependent resistance coefficient, lead to extrema\,(both maxima and minima) in the settling velocity profile; five different phases have been identified in the settling behavior of thin disks. A detailed theoretical investigation to establish the variation of the settlng velocity profile for small $Re$ and$Ri_v$ requires an integration of the coupled translational and orientational equations of motion, and this will be reported separately. It is worth noting one interesting feature in these experiments, however. The particles used in the experiments have a density that is greater than that of the lower denser layer of the non-linearly stratified ambient, and the resulting absence of a neutral buoyancy level renders these experiments closer to the ideal steady state scenario of a constant $U$, thereby pointing to the possible relevance of the intermediate orientation equilibria identified in figures \ref{fig:thresholdprolate} and \ref{fig:thresholdoblate}. Interestingly, in \cite{Mrokowska_20201}, the author observes thick disks to behave differently from thin ones. On entering the transition layer, these disks appear to rotate from an initial broadside-on configuration, attained in the upper layer, towards an intermediate inclined orientation, before rotating back onto a broadside-on orientation in the lower homogeneous layer. The persistence of the inclined orientation in the transition layer appears consistent with the prediction of equilibrium orientations in figure \ref{fig:thresholdoblate}. The ratio $Ri_v/Re^{\frac{3}{2}}$ equals $\frac{\gamma L^{\frac{3}{2}} \mu^{\frac{1}{2}}g}{U^{\frac{5}{2}}\rho_0^{\frac{3}{2}}}$ in terms of the underlying physical parameters. Further, using the scale $F/(\mu LX_A)$ for $U$, one obtains the ratio as $(\frac{3X_A}{4\pi})^{\frac{5}{2}}\frac{\gamma \mu^3}{(\rho_0g)^{\frac{3}{2}}(\Delta \rho)^{\frac{5}{2}} Lb^{\frac{5}{2}}}$. Both the thick and thin disks used in the experiments of \cite{Mrokowska_20201} correspond to $\kappa \ll 1$, implying that $X_A(\kappa) \approx X_A(0)$ in the expression for $Ri_v/Re^{\frac{3}{2}}$ above. It is the thickness $b$ that varies significantly in going from the thin to the thick disk in the experiments, and the $b^{-\frac{5}{2}}$ scaling of the above ratio implies that the thick disk will correspond to a significantly lower value of $Ri_v/Re^{\frac{3}{2}}$. Thus, it is possible for the thin disk to correspond to an $Ri_v/Re^{\frac{3}{2}}$  above the upper threshold, with the thick disk falling in between the two thresholds above; In this sense, our predictions again appear broadly consistent with the observations in \cite{Mrokowska_20201}.

\section{Conclusions and Future work} \label{sec:conclusions}

To summarize, in this effort, we present the first rigorous theoretical description of the orientation dynamics of anisotropic particles in a stably stratified ambient. In particular, the stratification-induced hydrodynamic torque, acting on an anistropic particle, has been calculated for the first time, and for large $Pe$ in particular, the torque is shown to rotate both prolate and oblate spheroids towards an edgewise orientation regardless of the spheroid aspect ratio. The theoretical predictions with regard to the transitions between broadside-on and edgewise settling, and with regard to the existence of intermediate inclined equilibrium orientations, appear broadly consistent with very recent experiments. Unfortunately, a detailed quantitative comparison appears out of reach at the moment, given that the particles used in the all of the experiments referred to in section \ref{sec:results} correspond to $Ri_v$ and $Re$ values of order unity and higher. We hope that future experiments will use smaller particles, in an attempt to access the regime of small $Re$ and $Ri_v$, and thereby validate the detailed predictions given here. It needs to be emphasized that the many of the smaller zooplankton, for typical values of the stratifcation corresponding to the oceanic pycnocline, correspond to the small $Re-Ri_v$ regime, and thus the theoretical framework given here is certainly relevant to natural settings\,(the oceanic realm in particular).

It is worth mentioning that the emphasis in the present manuscript has been on the large-$Pe$ analysis, in an attempt to explain the transition between broadside-on and edgewise settling observed in experiments all of which correspond to a salt-stratified ambient, and for the millemeter-sized particles used, therefore pertain to large $Pe$. It is of interest to extent the analysis here in a couple of obvious directions. For instance, scaling arguments discussed briefly towards the end of section \ref{sec:smallPe} highlight the possibility of an analogous broadside-on-edgewise transition even in the limit $Pe \ll 1$\,(specifically, $Ri_v^{\frac{3}{5}} \ll Pe \ll 1$). Further, the emergence of an $O(Ri_v^{\frac{2}{3}})$ torque for $Pe \gg Ri_v^{\frac{1}{3}}$ suggests that the large-$Pe$ analysis might have a wider applicability than originally intended; in that, the torque obtained in section \ref{sec:largePe} might already be valid for $Pe \gg Ri_v^{\frac{2}{3}}$ for $Ri_v \ll 1$. Finally, the opposing senses of rotation of oblate spheroids, with $\kappa < 0.17$ in the small-$Pe$\,($Pe \ll Ri_v^{\frac{3}{5}}$) and large-$Pe$ regimes point to a non-trivial dependence of the stratification-induced angular velocity as a function of $Pe$. These aspects will be taken up in future work. It is also of interest to move beyond orientation dynamics, towards a more detailed illustration of actual particle trajectories which requires an integration of the quasi-steady equations of motion for both translational and rotational degrees of freedom. This would enable a more detailed and fruitful comparsion with the experiments of \cite{Mrokowska_20201} and \cite{Mrokowska20202}. We expect some of the non-trivial signatures to be revealed in an analysis that might only incorporate a Stokesian drag for the positional dynamics, a valid leading order approximation for $Re, Ri_v \ll 1$. This will again be taken up in a future effort.

\nocite{*}
\section*{Acknowledgements}
Numerical computations reported here are mainly carried out by using the 'Nalanda-2' computational cluster available with JNCASR. The authors thank the institute for providing this facility.

\appendix
\section{Resistance functions and inertial torque}\label{appendixresistance}
The expressions for $F_I^p(\xi_0)$ and $F_I^o(\xi_0)$ defined in (\ref{eqn:prolate_inertialtorque}) and (\ref{eqn:oblate_inertialtorque}) are given in terms of the eccentricity of the spheroid ($e=1/\xi_0$) as
\begin{eqnarray}
F_I^p(\xi_0)=&&{-\pi e^{2} \left(420e+2240e^{3}+4249e^{5}-2152e^{7}\right)\over 315((e^{2}+1)\tanh^{-1}e-e)^{2}((1-3e^{2})\tanh^{-1}e-e)}\nonumber\\
+&&{\pi e^{2} \left(420+3360e^{2}+1890e^{4}-1470e^{6}\right)\tanh^{-1}e\over 315((e^{2}+1)\tanh^{-1}e-e)^{2}((1-3e^{2})\tanh^{-1}e-e)}\nonumber\\
-&&{\pi e^{2} \left(1260e-1995e^{3}+2730e^{5}-1995e^{7}\right)(\tanh^{-1}e)^{2}\over 315((e^{2}+1)\tanh^{-1}e-e)^{2}((1-3e^{2})\tanh^{-1}e-e)},\
\label{F_prolateinert}
\end{eqnarray}
and
\begin{eqnarray}
F_I^o(\xi_0)=&&{\pi e^{3}\sqrt{1-e^{2}}\left(-420+3500e^{2}-9989e^{4}+4757e^{6}\right)\over 315\sqrt{1-e^{2}}(-e\sqrt{1-e^{2}}+(1+2e^{2})\sin^{-1}e)(e\sqrt{1-e^{2}}+(2e^{2}-1)\sin^{-1}e)^{2}}\nonumber\\
+&&{210\pi e^{2}\left(2-24e^{2}+69e^{4}-67e^{6}+20e^{8}\right)\sin^{-1}e\over 315\sqrt{1-e^{2}}(-e\sqrt{1-e^{2}}+(1+2e^{2})\sin^{-1}e)(e\sqrt{1-e^{2}}+(2e^{2}-1)\sin^{-1}e)^{2}}\nonumber\\
+&&{105\pi e^{3}\left(12-17e^{2}+24e^{4}\right)(\sin^{-1}e)^{2}\over 315(-e\sqrt{1-e^{2}}+(1+2e^{2})\sin^{-1}e)(e\sqrt{1-e^{2}}+(2e^{2}-1)\sin^{-1}e)^{2}}.
\label{F_oblateinert}
\end{eqnarray}

The resistance functions $X_A$, $Y_A$ and $Y_C$ are expressed in terms of eccentricity as
\begin{align}
 X_A =\frac{16 \pi e^3}{3\left(2 e- (1+e^2)\log\left(\frac{1+e}{1-e}\right)\right)} 
\end{align}
\begin{align}
 Y_A =-\frac{32 \pi e^3}{3\left(2 e+ (3e^2-1)\log\left(\frac{1+e}{1-e}\right)\right)} 
\end{align}
\begin{align}
 Y_C =\frac{32 \pi e^3(e^2-2)}{3\left(-2 e+ (1+e^2)\log\left(\frac{1+e}{1-e}\right)\right)} 
\end{align}
for a prolate spheroid and
\begin{align}
 X_A=-\frac{8\pi e^3}{\left[e \sqrt{1-e^2}+(2 e^2-1)\cot^{-1}\left(\frac{\sqrt{1-e^2}}{e}\right)\right] }
\end{align}
\begin{align}
 Y_A=\frac{16\pi e^3}{\left[e \sqrt{1-e^2}-(1+2 e^2)\cot^{-1}\left(\frac{\sqrt{1-e^2}}{e}\right)\right] }
\end{align}
\begin{align}
 Y_C=\frac{16\pi e^3(e^2-2)}{3\left[e \sqrt{1-e^2}-(1-2 e^2)\cot^{-1}\left(\frac{\sqrt{1-e^2}}{e}\right)\right] }
\end{align}
for an oblate spheroid.
\bibliographystyle{jfm}
\bibliography{torque_anisotropic_Version12}
\end{document}